\documentclass[showpacs,floatfix]{revtex4}
\usepackage{amsmath}
\usepackage{amsfonts}
\usepackage{amssymb}
\usepackage{graphicx}
\usepackage{subfigure} 
\usepackage{bm} %bold math
\begin{document}

\title{Pulse Dynamics in a Chain of Granules With Friction}
\author{Alexandre Rosas and Katja Lindenberg}
\affiliation{
Department of Chemistry and Biochemistry,
and Institute for Nonlinear Science
University of California San Diego
La Jolla, CA 92093-0340}
\date{\today}

\begin{abstract}
We study the dynamics of a pulse in a chain of granules with friction.
We present theories for chains of cylindrical granules (Hertz potential
with exponent $n=2$) and of granules with other geometries ($n>2$). Our
results are supported via numerical simulations for cylindrical 
and for spherical granules ($n=5/2$). 
\end{abstract}
\maketitle

\section{Introduction}
\label{sec:introduction}
The propagation of pulses in granular materials has been a subject of
vigorous recent interest.  Seminal work on this subject was carried out
by Nesterenko~\cite{nesterenko} two decades ago in his analysis of
the propagation of
nonlinear compression pulses along a line of particles.  This early work
firmly established the nonlinear flavor of the problem: Nesterenko
showed that
under appropriate assumptions, among them the slow spatial variation of the
displacements of the particles, the equations of motion for 
granular particles could in most cases be approximated by a
continuous nonlinear partial
differential equation that admits a soliton solution (later shown to
actually be a solitary wave solution~\cite{mackay,senmanciu})
for a propagating perturbation in the
chain.  The recent revival of interest in the
subject~\cite{mackay,senmanciu,sinkovits,rogers,coste,sen98,naughton,hinch,hascoet,hascoet2,sen-book,hong,sen-pasi,wu,nakagawa}
has been triggered in part by a concern with
important technological applications such as the design of efficient
shock absorbers~\cite{sen-pasi}, the detection of buried
objects~\cite{rogers,coste,sen98,naughton}, and the fragmentation of
granular chains~\cite{hinch}.  The revival has
involved advances in the modeling, simulation, and solution of the
equations associated with important features of granular materials such as
their discreteness~\cite{mackay,senmanciu,hascoet,sen-pasi},
dimensionality~\cite{sen-pasi}, disorder~\cite{sen98,hascoet2,sen-pasi},
and loading provided
by gravitational forces~\cite{sinkovits,sen98,hong,sen-pasi,sen96,hascoet3}. 
The preponderance of the work has been numerical,
but important analytic advances have also been made, as well as experimental 
verifications of some of the theoretical
predictions~\cite{coste,nakagawa}.

The standard generic model potential between monodisperse elastic
granules that repel upon overlap according to the Hertz law is given
by~\cite{hertz,landau}
\begin{equation}
\begin{array}{l l l}
V(\delta_{k,k+1})&=a|\delta|^{n}_{k,k+1}, \qquad &\delta\leq 0,\\ \\
V(\delta_{k,k+1})&= 0, \qquad &\delta >0.
\end{array}
\label{eq:hertz}
\end{equation}
Here
\begin{equation}
\delta_{k,k+1} \equiv 2R - \left[ (z_{k+1} +y_{k+1}) - (z_k +y_k)
\right],
\end{equation}
\begin{equation}
a= \frac{2}{5D(Y,\sigma)}\left( \frac{R}{2}\right)^{1/2}, \qquad
D(Y,\sigma) = \frac{3}{2}\left( \frac{(1-\sigma^2)}{Y}\right),
\end{equation}
$Y$ and $\sigma$ denote Young's modulus and Poisson's ratio, $z_k$
denotes the initial equilibrium position of granule $k$ in the chain,
and
$y_k$ is the displacement of granule $k$ from this equilibrium position.
The geometric parameter $R$ is the radius if the particles are elastic
spheres.  More generally, $R$ is determined by the principal radii of
curvature of the surfaces at the point of contact~\cite{landau}.
The exponent $n$ is $5/2$ for spheres, it is $2$ for cylinders, and
in general depends on geometry.

The displacement of the $k$-th granule ($k=1,2, \ldots, L$) in the chain
from its equilibrium position in a frictional medium
is governed by the equation of motion
\begin{equation}
m \frac{\mathrm{d}^2{y}_k}{\mathrm{d} \tau^2} = -\tilde{\gamma}
\frac{\mathrm{d}{y}_k}{\mathrm{d}\tau} - a (y_k - y_{k+1})^{n-1}
\theta (y_k - y_{k+1}) + a (y_{k-1} - y_{k})^{n-1}
\theta (y_{k-1} - y_{k}).
\label{eq:motion}
\end{equation}
Here $\theta(y)$ is the Heavyside function, $\theta(y)=1$ for $y>0$, 
$\theta(y)=0$ for $y<0$, and $\theta(0)=1/2$. 
It ensures that the particles interact only
when
in contact. Note that for open boundary conditions the second term
on the right hand side of this equation
is absent for the last granule and the third term is absent for
the first, while for periodic boundary conditions $y_{k+L}=y_k$.
In terms of the rescaled variables and parameters
\begin{equation}
y_k = \left( \frac{m v_0^2}{a} \right) ^{1/n} x_k, \qquad
\tau = \frac{1}{v_0} \left( \frac{m v_0^2}{a} \right) ^{1/n} t, \qquad
\gamma =\frac{\tilde{\gamma}}{m v_0} \left( \frac{m v_0^2}{a} \right)
^{1/n},
\end{equation}
Eq.~(\ref{eq:motion}) can be rewritten as
\begin{equation}
\ddot{x}_k = -\gamma \dot{x}_k - (x_k - x_{k+1})^{n-1} \theta (x_k -
x_{k+1}) +  (x_{k-1} - x_{k})^{n-1} \theta (x_{k-1} -
x_{k}),\label{eq:motion_rescaled}
\end{equation}
where a dot denotes a derivative with respect to $t$.

Initially the granules are placed along a line so that they
just touch their neighbors in their equilibrium positions (no
precompression), and
all but one granule, granule $i$, are at rest. The velocity of
granule $i$ is $v_0$ (the impulse).  In the rescaled variables
the initial conditions become $ x_k(0) = \dot{x}_k(0) = 0$, 
$\forall k \neq i$, $ x_i(0) = 0 $, and $ \dot{x}_i(0) = 1 $.  In our
work we set $i=1$.

In the absence of the frictional contribution
$-\tilde{\gamma}\mathrm{d}{y}_k/\mathrm{d}\tau$ or
$-\gamma \dot{x}_k$,
when $n>2$ an initial impulse settles into a pulse that
becomes increasingly narrow with increasing $n$, and
propagates at a velocity that is essentially constant and determined
by $n$ and by the amplitude of the pulse. For $n=2$ the pulse spreads in
time and travels at a constant velocity independent of pulse amplitude.
In the latter case there is considerable backscattering that leads to
backward motion of all the granules behind the pulse, whereas the
backscattering is minimal for $n>2$~\cite{hinch}.  The pulse is
a completely conservative solitary wave in the limit $n\rightarrow\infty$.

{\em Our interest lies in ascertaining the effects of the frictional
contributions on these results.}  
Although friction is on occasion mentioned in theoretical and numerical
work, it is
usually mentioned as an element that is neglected or omitted.  However, its
presence and importance in experiments is
recognized as inevitable~\cite{nakagawa}.  Note that 
frictional effects may arise not only from the forces between the granular
chain and the surrounding medium but also from the conversion of
translational motion to other degrees of freedom (e.g.
rotation)~\cite{nakagawa}.
In an earlier paper we analyzed in
detail the effects of frictional forces on the dynamics of two
granules, specifically the way in which forward and backward motion of the
granules is affected by these forces~\cite{we}.  Herein we extend that
work to a chain of granules.

Since our approach will in general follow established methods,
it is useful to lay out at the outset an overview of the principal
approaches that have been implemented in the study of pulse dynamics 
in frictionless chains, and the various features that determine these
dynamics.  
Theoretical studies of pulse dynamics in frictionless chains have
mainly relied on three
approaches:

(1) Numerical solution of the equations of
motion~\cite{sinkovits,sen98,hinch,hascoet,hascoet2,sen-pasi,hascoet3};

(2) Continuum
approximations to the equations of motion followed by exact or approximate
solutions of these approximate
equations~\cite{nesterenko,hinch,hascoet3}; and

(3) Phenomenology about
properties of pairwise
(or at times three-body) collisions together with the
assumption that pulses are sufficiently narrow to be principally
determined by these properties~\cite{wu,mohan}.  

From these studies it is clear that there three features determine
the dynamics in these chains:

(1) The power $n$ in the potential;

(2) The absence of a restoring force; and

(3) The discreteness of the system.

While each of these leads to essential aspects of the dynamics, the
literature is not always entirely clear on which feature is determinant
in a particular behavior, nor is it always
clear which of these features is being approximated or even ignored.
One example is the equivocal connection between the dynamics
of granular systems and systems
with power law interaction potentials that {\em include} a restoring
force. For instance, there is an extensive literature on the Fermi-Pasta-Ulam
(FPU) chain with purely nonlinear interactions of the form
appearing in Eq.~(\ref{eq:motion_rescaled}) with $n>2$ but without
the Heavyside theta functions~\cite{FPU}.  Highly localized pulses 
propagate in these systems~\cite{sen-pasi,ourpulse}, but their
relation to the localized solutions in granular systems is by no means
clear.  This ambiguity is exacerbated by the fact that discrete
systems are frequently approximated
by continuum equations. While the pulse-like (soliton or solitary wave)
solutions that emerge from these approximations are assumed (and even shown
numerically) to describe certain aspects of 
granular system, the continuum
approximations never explicitly respect the absence of a restoring
force, i.e., they are more clearly justifiable for FPU-like systems.  
And so with these solutions in hand, it is not clear what consequences
of the absence of a restoring force and of the discreteness of the
system have been lost. In some sense opposite to the continuum
approximations are two-body or three-body phenomenologies.  These
provide considerable qualitative insight into some consequences of
discreteness, but are quantitatively not sufficiently accurate because
the true dynamics extend over more than two or three granules.
In this paper, even while focusing on the effects of friction, we
make an attempt to provide some clarity on these issues.

Numerical simulations are of course extremely helpful and ultimately an
accurate test of theory.  Note that the rescaled system 
Eq.~(\ref{eq:motion_rescaled}) has no free parameters in the absence of
friction and only a single parameter when there is friction, and therefore
numerical characterization is particularly straightforward in this
particular system.  However, this universality is lost with any of a
number of modifications that might be interesting and have been considered
recently such as, for example, tapering of the masses in the
chain~\cite{senmanciu,wu,nakagawa} and mass~\cite{sen98,hascoet2} or
frictional~\cite{we} disorder.  Consequently, it is desirable to have a
strong analytic backdrop.

The paper is organized as follows.  Section~\ref{sec:cylinder} deals
with a chain of cylindrical granules.  First we review existing results
for frictionless granules, discussing them in the context of the issues
and uncertainties noted earlier.  We then extend the theory to granules
subject to friction, and present numerical simulations in support of
these results.  In Sec.~\ref{sec:sphere} we follow the same presentation
sequence for granules with $n>2$, with special attention to spherical
granules in our numerical simulations.  A summary of results and of future
directions is presented in Sec.~\ref{sec:conclusion}.

\section{Cylinders ($n=2$)
\label{sec:cylinder}}

Cylindrical granules have provided a theoretical testbed for dynamics in
granular chains because the exponent $n=2$ leads to analytic
manageability not available for other potentials.  Although 
sometimes called the ``harmonic'' case, it should be remembered that
the Hertz potential is quite different because there is no restoring force,
that is, the cylindrical Hertz potential is half of a harmonic potential.
The derivative of the force law therefore changes discontinuously
between extension and compression.
This leads to considerable differences in the chain dynamics. 
Nevertheless, some aspects of
the cylindrical chain dynamics can be inferred from those of a harmonic
chain, and this can be exploited to great advantage in the analysis.

If one were to ignore the absence of a restoring force and implement the
simplest continuum approximation to Eq.~(\ref{eq:motion_rescaled}) in
the absence of friction, the result would be a simple wave equation
with diffusive coupling
whose solutions do not represent the observed behavior of the $n=2$
chain.  In reality, an initial impulse in the chain described by 
Eq.~(\ref{eq:motion_rescaled}) in the absence of friction moves as a
spreading pulse. Although the pulse spreads and sheds some energy, the
waveform can nevertheless be clearly identified as a pulse~\cite{hinch}. 
Its maximum $k_{max}(t)$ travels forward with a constant unit velocity
and a displacement amplitude that increases as $t^{1/6}$.  The pulse
spreads in time as $t^{1/3}$, {\em more slowly} than it would in a system
with diffusive coupling.  It is interesting to
understand which of these features are due to the
discrete nature of the chain, and which are due to the absence of a
restoring force.  Further, once these questions have been answered,
it is interesting to explore what happens to these features in the
presence of friction.

Our presentation of the chain of cylindrical granules consists of three
parts.  First we review the results of Hinch and
Saint-Jean~(HSJ)~\cite{hinch}
for a frictionless chain, and we recast the problem in a way that
clarifies some of the approximations made in that work
and pinpoints the sources of differences between the $n=2$
chain and a purely diffusive coupling.  We supplement this review with
our analytic results that reproduce some of their purely numerical
ones.  Then we modify this analysis
to include the effects of weak
hydrodynamic friction on the granules.  Finally, we complement this
analytic (and necessarily approximate) treatment with a comparison with
numerical simulation results for the frictional chain of cylindrical
granules.

\subsection{Frictionless Granules - Theory}
The HSJ theory is based on the
following approximations implemented consecutively and independently. 

1. The solution is assumed to be described by a traveling pulse
of constant form which
propagates at constant unit speed and has an amplitude $b$ and width
$\lambda$ that vary slowly with time:
\begin{equation}
x(k,t)= b(t) f\left(\frac{k-t}{\lambda(t)}\right).
\label{eq:assume}
\end{equation}

2. The pulse is assumed to retain almost all of its initial energy.  HSJ
show that assumption 1. leads to equipartition of this energy between
potential and kinetic forms, as should be the case for a harmonic
potential.

These two assumptions are sufficient
to lead to the conclusion that $\lambda\propto b^2$.

3. A continuum approximation is implemented that takes into account some
discreteness effects.  In the lowest order diffusive coupling
approximation one would set
the difference $x_{k+1}+x_{k-1}-2x_k\approx \partial^2x(k)/\partial k^2$.
Retention of the next term in a Taylor series expansion of
$x_{k\pm 1}$ about $x_k$ leads to the continuum approximation that
incorporates some of the effects of discreteness:
\begin{equation}
\frac{\partial^2 x(k,t)}{\partial t^2} =
\frac{\partial^2 x(k,t)}{\partial k^2} + \frac{1}{12}
\frac{\partial^4 x(k,t)}{\partial k^4}.
\label{eq:wavehinch}
\end{equation} 
{\em Note that this expansion includes a restoring force.}
A transformation to a
moving frame with unit propagation velocity, i.e., a change of variables from
$k$ and $t$ to $\nu=k-t$ and $t$, transforms this equation to
\begin{equation}
\frac{\partial^2 x}{\partial t^2} -
\frac{\partial^2 x}{\partial t \partial \nu} = \frac{1}{12}
\frac{\partial^4 x}{\partial \nu^4}.
\label{eq:wavehinchmove}
\end{equation} 

Regardless of the form of $b(t)$ or of the function $f$, this is sufficient
to establish that the solution Eq.~(\ref{eq:assume}) is consistent with
this equation only if $\lambda\sim t^{1/3}$, and also that the next
term in the Taylor series expansion is unimportant for this result. 
{\em The width of the pulse is
therefore governed by the first manifestation of discreteness.} One
assumes that the absence of a
restoring force does not affect this result because the granules within the
pulse are in fact overlapping most of the time and hence subject
mainly to the repulsive portion of the potential.  Note that the continuum
approximation together with the conservation of energy are then sufficient
to arrive at the conclusion that $b(t)\sim t^{1/6}$, i.e., that the pulse
amplitude actually grows with time.

4. It is not yet clear that (\ref{eq:assume}) is actually compatible with
(\ref{eq:wavehinch}) until one determines the function $f$.  Substitution
of (\ref{eq:assume}) into (\ref{eq:wavehinch}) and retention of leading terms
in $t$ leads to
the equation for $f(\xi)$, 
\begin{equation}
f^{''''} -8\xi f^{''}-4f^{'}=0.
\label{eq:fz2}
\end{equation}
This equation has four solutions. The one consistent with the
requirement that $f(\xi)$ decays for large $\xi$ (i.e., ahead of the
wave) and consistent with the assumption of energy conservation by
the pulse is~\cite{as}
\begin{equation}
f(\xi)= N \int_\xi^\infty Ai^2(2^{1/3}y){\mathrm d}y
\label{eq:airy}
\end{equation}
where $Ai(z)$ is an Airy function and 
\begin{equation}
N= \left[ \frac{E_p}{\int_{\xi_0}^\infty Ai^4(2^{1/3}y){\mathrm
d}y}\right]^{1/2} = 3.533 \sqrt{E_p}.
\label{eq:normalization}
\end{equation}
Here $E_p$ is the pulse energy and $2^{1/3}\xi_0$ is the
first zero of $Ai(z)$.  

These features describe the traveling displacement pulse of increasing width 
and amplitude quite accurately, as shown by the numerical
simulations in~\cite{hinch}. 
We note that the total energy of the system with the initial unit velocity
impulse at one granule is $1/2$. The numerical results of HSJ lead to an
asymptotic pulse energy of $E_p=0.48$, that is, $96.2\%$ of the energy
resides in the pulse. Below we calculate the pulse energy analytically.

In addition to the forward traveling pulse, HSJ also investigate the
momentum of the particles ejected backward as the pulse goes by.  That
particles must be ejected is a consequence of the conservation of momentum:
the traveling pulse of constant energy and increasing width and
displacement amplitude
carries increasing forward momentum, which must be balanced by the backward
momentum of the ejected particles.  One then arrives at the next item on the
list of assumptions:

5. The absence of a restoring force is explicitly recognized in the
calculation of the momentum of the ejected particles, which simply
keep traveling backward
with the momentum they acquire at the moment of separation from the pulse.
Equating the rate of change of the momentum of the forward pulse at time
$t$ to that of the particle ejected at that time leads to the conclusion
that the backward momentum of the $n^{th}$ particle is $\dot{x}_n =-c
t^{-5/6}=-c n^{-5/6}$, the latter equality arising from the
unit speed of pulse propagation.  The numerical simulations of HSJ
lead to the value $c=0.158$, a value that we obtain analytically.

To obtain analytic results for the pulse energy $E_p$ and the
constant of proportionality $c$, we note that the forward momentum of
the propagating pulse is 
\begin{eqnarray}
P&=&\sum_{\dot{x}(k,t)>0} \dot{x}(k,t) = \sum_{\dot{x}(k,t)>0}
\frac{b(t)}{\lambda(t)}f'(\xi)\nonumber\\
&=& b(t) \int_{\xi_0}^\infty f'(\xi) {\mathrm
d}\xi
= \widetilde{N} b(t)
\end{eqnarray}
where
\begin{equation}
\widetilde{N} = N \int_{\xi_0}^\infty Ai^2(2^{1/3}\xi) {\mathrm d} \xi =
1.379\sqrt{E_p}.
\label{eq:ntilde}
\end{equation}
The rate of change of the forward momentum then is
$\dot{P} =\widetilde{N}\dot{b}(t)$.
Since the pulse velocity is unity, this is
the momentum transferred to the last particle as it is ejected:
\begin{equation}
v_b = - \dot{P} 
= -\widetilde{N} \dot{b}(t) 
= -\frac{\widetilde{N}}{6} t^{-5/6}  = -\frac{\widetilde{N}}{6}
n^{-5/6}. 
\label{eq:vsubb}
\end{equation}
The backscattered energy then is
\begin{equation}
E_b = \frac{1}{2} \left(\frac{\widetilde{N}}{6}\right)^2 \sum_{k=1}^\infty
k^{-5/3} = 0.056 E_p
\end{equation}
and for the total energy we obtain
\begin{equation}
E=E_p+E_b = 1.056E_p,
\end{equation}
from which it immediately follows that asymptotically the energy in the
pulse is 
\begin{equation}
E_p=0.947 E.
\end{equation}
Thus, 94.7\% of the energy resides in the pulse (to be compared to the
HSJ value of $96.2\%$ obtained from simulations) and the remainder is
backscattered.  With the initial condition $E=1/2$ used in all
simulations we thus have for the constants defined earlier $N=2.431$ and
$\widetilde{N}=0.949$. The constant $c$ in HSJ is
$c=\widetilde{N}/6=0.158$, exactly as they obtained from numerical
simulations.

This essentially completes the solution.  The summary description is
then that an initial velocity impulse propagates forward at unit speed,
with a width that grows as $t^{1/3}$ and a displacement amplitude that
grows as $t^{1/6}$.  This pulse carries almost all of the initial
energy and its momentum increases. 
This increase in momentum is compensated by granules that
are ejected backward as the pulse passes. The speed of the ejected
granules decreases, the $n^{th}$ granule being ejected with a speed
proportional to $n^{-5/6}$.

Further insights can be gained by viewing this problem a bit differently.
Suppose that we do not implement a continuum approximation at all, but
instead focus on the full discrete equation
\begin{equation}
\ddot{x}_k = x_{k+1}+x_{k-1}-2x_k.
\label{eq:discrete}
\end{equation}
Like the continuum equation, this of course includes restoring forces.
This equation can be solved exactly~\cite{ourpulse}, 
\begin{equation}
x_k(t)=\int_0^t J_{2k-2}(2t')dt',
\label{eq:bessel}
\end{equation}
where $J_n(z)$ is the Bessel function of the first kind~\cite{as}.  
The solution is a moving spreading pulse along with oscillatory 
displacements that are left in its wake (precisely
because there are restoring forces).  The peak and width of the pulse
can be obtained from the properties of the Bessel functions, in
particular from knowledge of their first two zeroes and the first
maximum~\cite{royal}.  The maximum of the pulse occurs at
$k_{max}(t)=t+\mathcal{O}(t^{1/3})$, so the pulse velocity is unity.
Its width increases as
$t^{1/3}$.  In these features the solution is similar to the one assumed in
Eq.~(\ref{eq:assume}).
However, the amplitude of the displacement pulse does not grow but
is instead constant
in time.  (If we write the Bessel function
solution in the form (\ref{eq:assume}) but with $b(t)=const$ independent of
$t$, we find from substitution into Eq.~(\ref{eq:wavehinch})
that $f$ satisfies the equation  
$f^{''''} -8z f^{''}-8f^{'}=0$.)
The pulse energy according to this description is not constant but
instead decreases as $t^{-1/3}$. The energy that is lost goes into the
oscillatory
displacements in the wake of the harmonic pulse caused by the restoring
forces.  The {\em additional} and appropriate assumption of {\em energy
conservation} in HSJ adds this lost energy back into the pulse 
without affecting
its spreading rate or velocity.  {\em This indicates that}
(in the $n=2$ problem)
{\em energy conservation is an appropriate additional assumption in lieu
of the absence of a restoring force}.

In summary: the fact that the solution of the $n=2$ Hertz problem is a
spreading pulse of unit velocity and of width that increases as $t^{1/3}$
is purely a consequence of the discreteness of the system and not dependent
on the presence or absence of restoring forces; it is a feature of a
discrete harmonic system.  The (near) conservation of energy in the pulse
is {\em not} a feature of a
harmonic system and must be included as an {\em additional} assumption.
Conservation of energy must be
implemented not only to describe all the features of the traveling pulse
but also to describe the backward momentum of the particles ejected
as the pulse moves along.  Extensive numerical results quantitatively
supporting the features just described can be found in~\cite{hinch}.

\subsection{Granules With Friction - Theory}

Here we generalize the previous theories to a chain in which the
granules experience friction [cf. Eq.~(\ref{eq:motion_rescaled})].
We generalize the HSJ theory to this case
and also use the solution of the damped harmonic chain to
complement these results.  

1. The solution is assumed to be a traveling pulse of constant shape which
propagates at unit speed, has a width $\lambda$ that varies slowly in
time, and an amplitude that {\em aside from an exponential decay due to the
friction} also varies slowly with time:
\begin{equation}
x(k,t)=  e^{-\gamma t/2}b(t)f\left(\frac{k-t}{\lambda(t)}\right).
\label{eq:assumef}
\end{equation}
We retain only the $\gamma$-independent portions of the slowly varying
functions $\lambda$ and $b$. This is thus a ``lowest order'' ansatz.
Indeed, we will show later that corrections to this lowest order
ansatz seem to be of $\mathcal{O}(\gamma^3)$.
The pulse energy then decays exponentially as $e^{-\gamma t}$.
Explicitly, the kinetic energy is obtained as follows.
Retaining only leading orders in time, the pulse
velocity that follows from Eq.~(\ref{eq:assumef}) is
\begin{equation}
\dot{x}(k,t) =
-e^{-\gamma t/2}\frac{b(t)}{\lambda(t)} f'\left(\frac{k-t}{\lambda(t)}\right)
+\mathcal{O}(\gamma t).
\label{eq:velocityf}
\end{equation}
Therefore, the kinetic energy to $\mathcal{O}(\gamma t)$ is
\begin{eqnarray}
K(t) &=& e^{-\gamma t}\frac{1}{2}\sum_k \left [\frac{b(t)}{\lambda(t)}
f'\left(\frac{k-t}{\lambda(t)}\right) \right ]^2 \simeq
e^{-\gamma t}\frac{1}{2}\frac{b^2(t)}{\lambda(t)} \int_{\xi_0}^\infty
\left[ f'(\xi)\right]^2 {\mathrm d}\xi
\nonumber\\
&=& \frac{E_p}{2} e^{-\gamma t}
\frac{b^2(t)}{\lambda(t)},
\end{eqnarray}
where $E_p$ is the undamped pulse energy.
On the other hand, the potential energy may be written as
\begin{equation}
U(t) = \frac{1}{2}\sum_k (x_{k+1}-x_k)^2 \simeq \frac{1}{2}\sum_k
\left[\frac{\partial
x(k,t)}{\partial k}\right]^2,
\end{equation}
which also leads to 
\begin{equation}
U(t) = \frac{E_p}{2} e^{-\gamma t}
\frac{b^2(t)}{\lambda(t)} 
\end{equation}
to $\mathcal{O}(\gamma t)$.  
The total pulse energy to this order then is 
\begin{equation}
\label{eq:energy}
E_p(t) = E_p e^{-\gamma t}\frac{b^2(t)}{\lambda(t)}.
\end{equation}

2. We assume that almost all of the energy resides in the decaying
pulse.  

These results and assumptions are thus again sufficient to conclude that
$\lambda\propto b^2$.  It then follows from Eq.~(\ref{eq:energy}) that the
only effect of friction to this order is thus the overall exponential
decay of the energy.

3. As before, a continuum approximation is implemented that takes into
account some discreteness effects and of course the frictional contribution 
(while still including the nonexistent restoring force, as before).
For this purpose it is convenient to implement the change of variables 
$z(k,t)=e^{\gamma t/2}x(k,t)$.
Retention of the next term in a Taylor expansion of
$z_{k\pm 1}$ about $z_k$ beyond the purely diffusive approximation
leads to 
\begin{equation}
\frac{\partial^2 z(k,t)}{\partial t^2} =
\frac{\partial^2 z(k,t)}{\partial k^2} + \frac{1}{12}
\frac{\partial^4 z(k,t)}{\partial k^4}+\frac{\gamma^2}{4}z(k,t).
\label{eq:wavehinchf}
\end{equation}
The associated moving frame equation ($\nu=k-t$) is 
\begin{equation}
\frac{\partial^2 z }{\partial t^2} -
\frac{\partial^2 z}{\partial t \partial \nu} = \frac{1}{12}
\frac{\partial^4 z}{\partial \nu^4}+\frac{\gamma^2}{4}z.
\label{eq:wavehinchfmoving}
\end{equation}
With the ansatz
\begin{equation}
z(k,t)= b(t) f\left(\frac{k-t}{\lambda(t)}\right)
\label{eq:assumey}
\end{equation}
associated with (\ref{eq:assumef}) this is again sufficient to establish
that the pulse widens as $\lambda\sim t^{1/3}$. 
Corrections begin to set in when the last term in
Eq.~(\ref{eq:wavehinchfmoving}) becomes important, i.e.,
for times $t\gtrsim \gamma^{-3/2}$.  For small $\gamma$ this is a time 
much longer than the entire lifetime of the pulse, which is of
$\mathcal{O}(\gamma^{-1})$ due to the overall exponential decay.  As
before, we are again led to the conclusion that $b(t)\sim t^{1/6}$.

4. The envelope function is again the solution of Eq.~(\ref{eq:fz2}) with
corrections of $\mathcal{O}(\gamma^2)$.

5. As before, as the pulse travels forward granules are ejected backwards.
In the frictionless problem these granules continue moving backward
with the same velocity forever after ejection because they are simply
freely moving granules.  In the problem with friction
these granules progressively slow down because they, too, are damped, but it is
of interest to calculate their momentum at the moment of ejection from
the pulse.  Indeed, we now show that in the presence of damping the
backward ejection momentum is {\em greater} than in the undamped chain, a
result consistent with that found in a two-granule system~\cite{we}.  

The forward momentum of the propagating pulse to $\mathcal{O}(\gamma t)$ is
\begin{eqnarray}
P&=&\sum_{\dot{x}(k,t)>0} \dot{x}(k,t) = \sum_{\dot{x}(k,t)>0}
e^{-\gamma t/2}\frac{b(t)}{\lambda(t)}f'(\xi)\nonumber\\
&=& e^{-\gamma t/2}b(t) \int_{\xi_0}^\infty f'(\xi) {\mathrm d}\xi
= \widetilde{N} e^{-\gamma t/2} b(t),
\end{eqnarray}
where $\widetilde{N}$ is given in Eq.~(\ref{eq:ntilde}).  
The rate rate of change of the forward momentum is
\begin{equation} 
\dot{P} =\widetilde{N}\dot{b}(t) e^{-\gamma t/2} \left( 1-\frac{\gamma
b(t)}{2\dot{b}(t)}\right).
\end{equation}
The last particle in the pulse loses momentum $\delta$
through dissipation as the
pulse moves across it.  This momentum loss is the difference in the
momentum of this particle when it is in the middle of the pulse (i.e.,
when it is maximally compressed, at which point its momentum is zero)
and when it is at the end of the pulse and about to be ejected
(at which point its compression is zero):
\begin{equation}
\delta = -\gamma \Delta x =-\gamma x_{max} = -\gamma \widetilde{N}
e^{-\gamma t/2} b(t).
\label{eq:loss}
\end{equation}
Since the pulse velocity is still unity, the
momentum transferred to the last particle as it is ejected is 
\begin{eqnarray}
v_b &=& - \dot{P} +\delta
= -\widetilde{N} e^{-\gamma t/2} \dot{b}(t) \left(
1-\frac{\gamma b(t)}{2\dot{b}(t)} +\frac{\gamma b(t)}{\dot{b}(t)}\right)
\nonumber\\
&=& -\frac{\widetilde{N}}{6} e^{-\gamma t/2} t^{-5/6} \left( 1+3\gamma
t\right).
\label{eq:vfsubb}
\end{eqnarray}
To facilitate comparison of this result with that of the frictionless
chain we expand the exponential and retain terms of $\mathcal{O}(\gamma t)$: 
\begin{equation}
v_b = -\frac{\widetilde{N}}{6}t^{-5/6} \left(1+\frac{5}{2}\gamma t\right).
\label{eq:larger}
\end{equation}
{\em This confirms that the
backward ejection momentum is indeed greater than in the undamped chain.}

The forward moving pulse of the $n=2$ Hertz problem with small friction is 
essentially identical in shape to that of the frictionless problem except
for an overall exponential decay. The pulse travels at unit speed and its
width increases as $t^{1/3}$, these two features again being a consequence
of discreteness and essentially unaffected by friction.  The assumption
that the pulse energy decreases only
because of the friction is an additional assumption.  This feature must 
again be implemented separately to describe the amplitude of the
traveling pulse
correctly, and also to describe the backward momentum of the granules
ejected as the pulse moves along.  

An analysis of the full discrete equation (with restoring forces) 
starts from the linear equation
\begin{equation}
\ddot{x}_k = x_{k+1}+x_{k-1}-2x_k -\gamma\dot{x}_k.
\label{eq:discretef}
\end{equation}
The change of variables $z_k(t)=e^{\gamma t/2}x_k(t)$ immediately leads
to 
\begin{equation}
\ddot{z}_k = z_{k+1}+z_{k-1}-2z_k +\frac{\gamma^2}{4}z_k
\label{eq:discretef2}
\end{equation}
and consequently to the solution
\begin{equation}
x_k(t)=e^{-\gamma t/2}\left[\int_0^t J_{2k-2}(2t')dt'+
\mathcal{O}(\gamma^2)\right].
\label{eq:besself}
\end{equation}
The solution is a moving spreading decaying pulse along with decaying
oscillatory displacements.  As before, the peak of the pulse occurs at
$k_{max}(t)=t+\mathcal{O}(t^{1/3})$, its width increases as $t^{1/3}$,
and the amplitude of the pulse decays exponentially.

%1
\begin{figure}
\vspace*{0.8cm}
\begin{center}
\includegraphics[width=8cm]{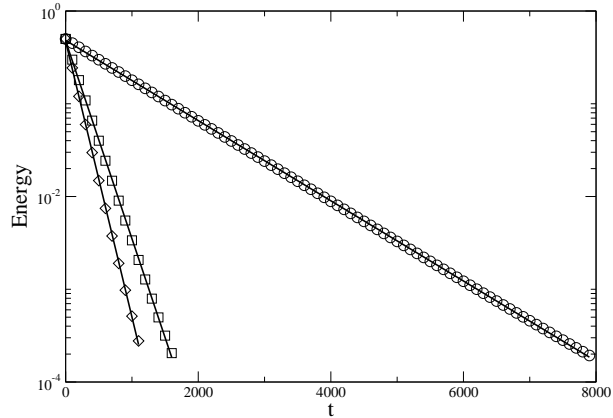}
\caption{
Total energy decay as a function of time. The energy decays
exponentially as $e^{-\gamma t}$ (lines). The symbols are the numerical
simulations: circles for $\gamma=0.001$, squares for $\gamma=0.005$,
diamonds for $\gamma=0.007$.
\label{fig:energy}}
\end{center}
\end{figure} 

In summary: the fact that the solution of the $n=2$ Hertz problem
with friction is an exponentially decaying
spreading pulse of unit velocity and of width that increases as $t^{1/3}$
is purely a consequence of the discreteness of the system and not dependent
on the presence or absence of restoring forces; it is a feature of a
discrete harmonic system.  The assumption that the pulse retains almost
all of the system energy (which decays exponentially)
is again {\em not} a feature of a harmonic system and must be
included as an {\em additional} assumption.
This conservation feature must be
implemented not only to describe correctly all the features of the
traveling pulse
but also to describe the backward momentum of the particles ejected
as the pulse moves along.  This backward momentum is greater in the chain
with friction than in the frictionless system.

\subsection{Numerical Simulations}

In this section we check the accuracy of our analytic results for
the $n=2$ frictional chain through numerical simulations. 
First, we note that
the total energy of the system decreases exponentially, as 
seen in Fig.~\ref{fig:energy} for various values of the friction
coefficient.  
For times $t\lesssim 10 \gamma^{-1}$, aside from the overall exponential
decay of the energy the system behavior is indeed the same as that of the
frictionless system in that it broadens as $t^{1/3}$ and propagates at
unit velocity. We find that the pulse ceases to exist at a time of order
$t\sim 10 \gamma^{-1}$.  More specifically, we find that at a time
$\sim 8.6\gamma^{-1}$ the pulse energy decreases abruptly (more rapidly
than exponential), and at a time $\sim 15.7\gamma^{-1}$ the
backscattered energy becomes greater than the pulse energy.

%2a,b,c,d
\begin{figure}
\begin{center}
\subfigure[$ \gamma = 0.000 $]{\includegraphics[width=6cm]{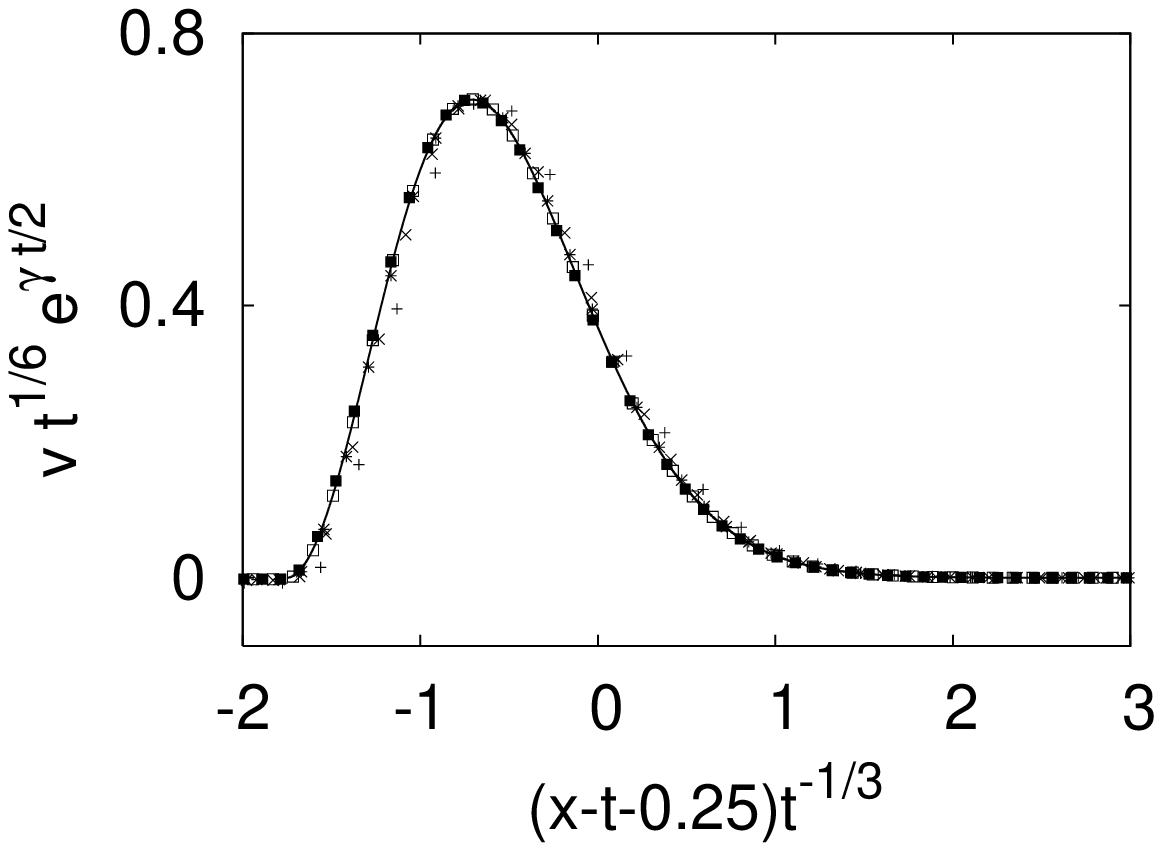}}
\subfigure[$ \gamma = 0.001 $]{\includegraphics[width=6cm]{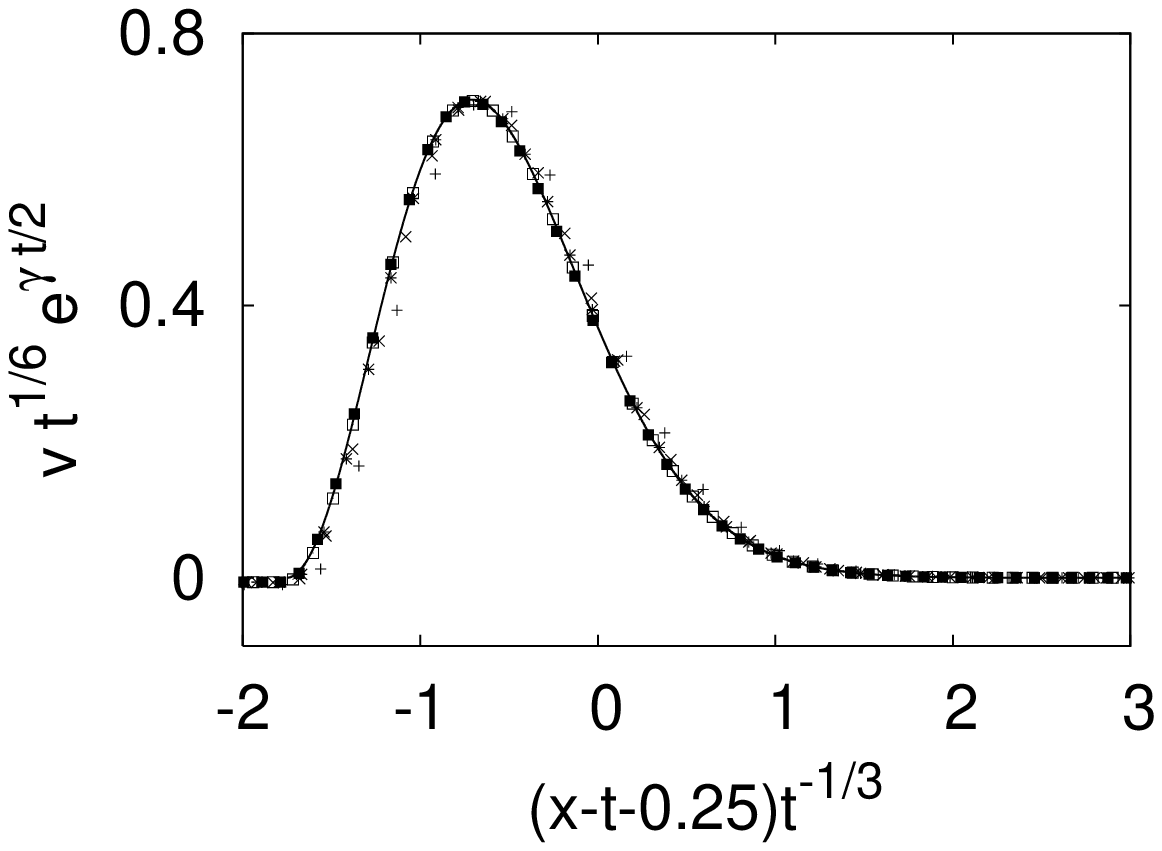}}
\subfigure[$ \gamma = 0.010 $]{\includegraphics[width=6cm]{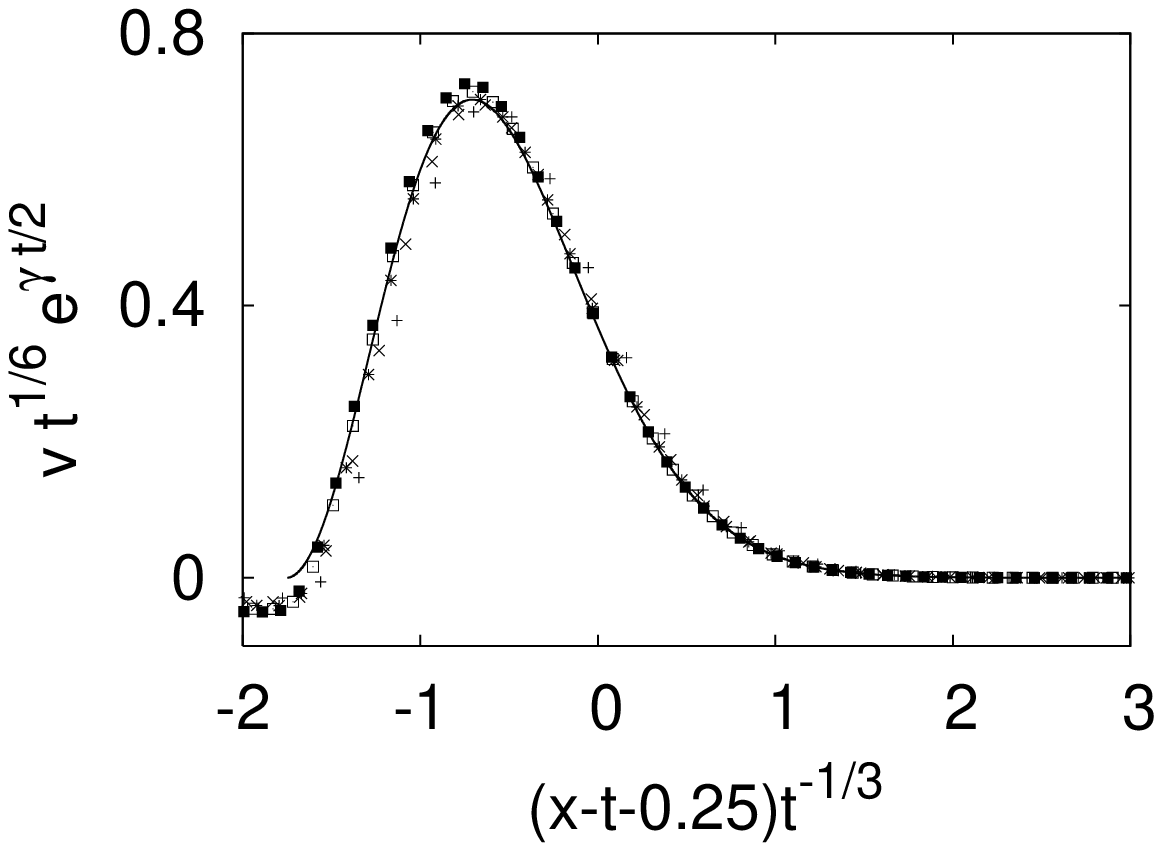}}
\subfigure[$ \gamma = 0.020 $]{\includegraphics[width=6cm]{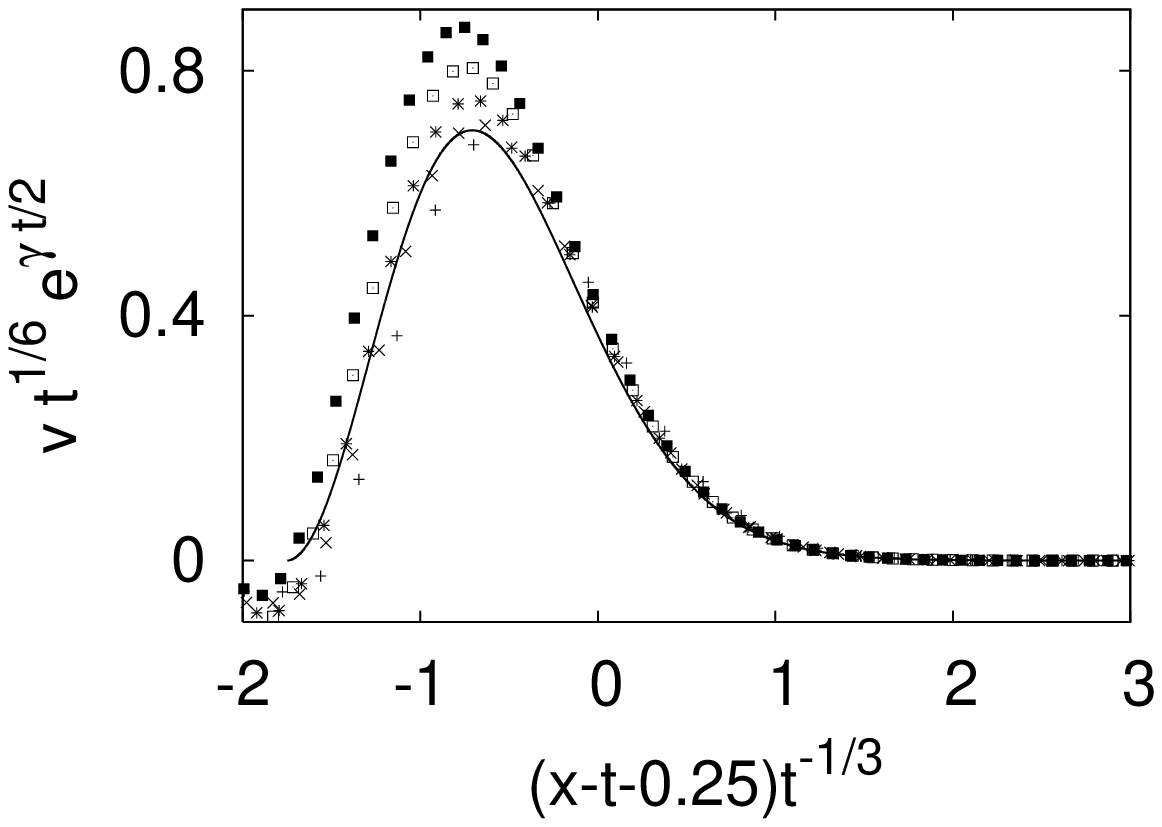}}
\end{center}
\caption{ Chain of cylindrical
granules: scaled velocity pulse
as a function of scaled position for different friction coefficients.
\label{fig:scaling}} 
\end{figure}

We must check our prediction for the self-similar impulse wave
$f(\xi)$ propagating along the chain of particles.  We follow HSJ and plot
the scaled velocity as a function of the scaled position for different
times in Fig.~\ref{fig:scaling}.  From Eq.~(\ref{eq:velocityf}) the
appropriate scaled
velocity to leading order in the friction is $f'(\xi)=\dot{x}\lambda e^{\gamma
t/2}/b \propto \dot{x} t^{1/6}e^{\gamma t/2}$, and this is the ordinate in
the panels. The abcissa is $\xi=(k-t-0.25)/\lambda \propto (k-t-0.25)t^{-1/3}$.
The shift $0.25$, which also occurs in the frictionless case,
is carefully explained and derived in HSJ and comes about
because the scaling function solution has velocities in the propagating
impulse wave of $\mathcal{O}(t^{-1/6})$ (here modified by the exponential
friction factor) while the rebound velocity of particles is of
$\mathcal{O}(t^{-5/6})$ (again modified by the exponential factor).  This
indicates that there must be a correction of $\mathcal{O}(t^{-2/3})$ to the
leading-order term, and this correction appears as a shift in the scaling.
The solid curve is $f'$ as obtained from Eq.~(\ref{eq:airy}), while the points
are the results of our numerical simulations.  The agreement is very good
provided the damping is small, but serious deviations begin to set in
with increasing damping, as seen in panel (d).

%3a,b,c,d
\begin{figure}
\begin{center}
\subfigure[$ \gamma = 0.000 $]{\includegraphics[width=6cm]{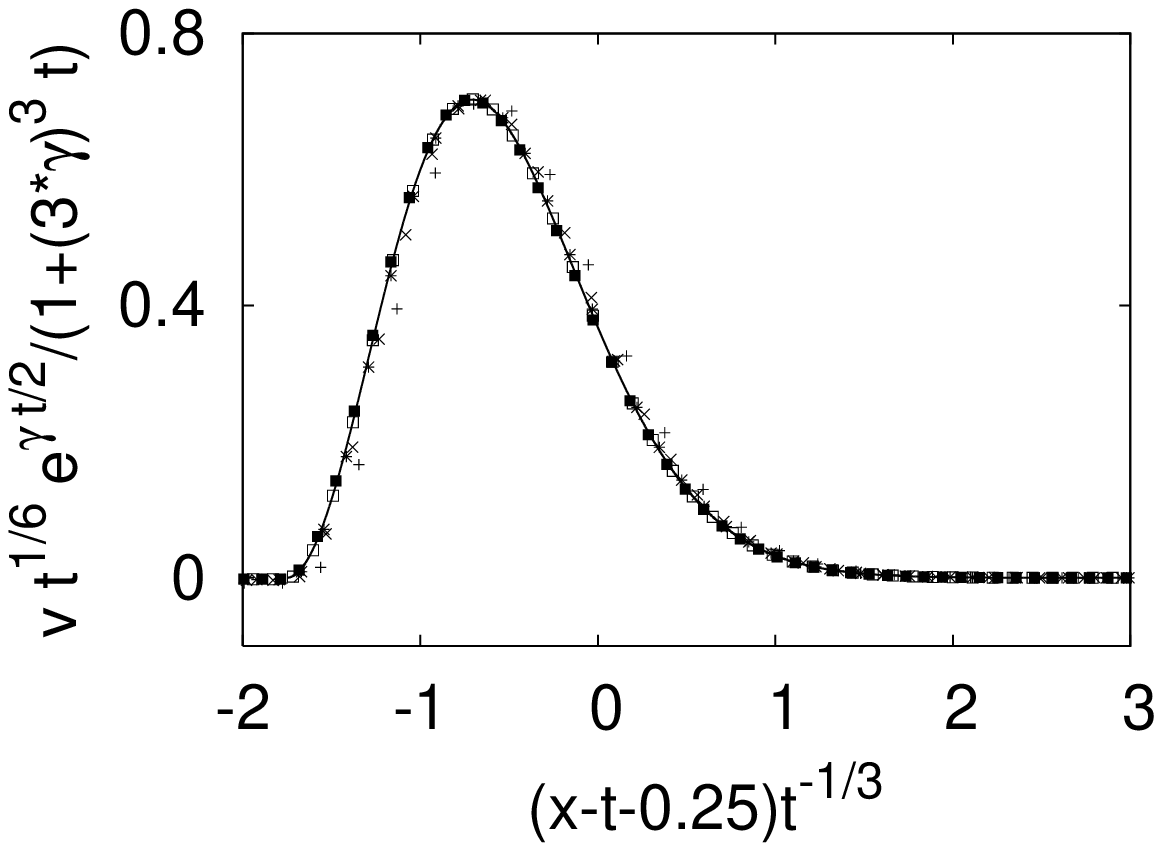}}
\subfigure[$ \gamma = 0.001 $]{\includegraphics[width=6cm]{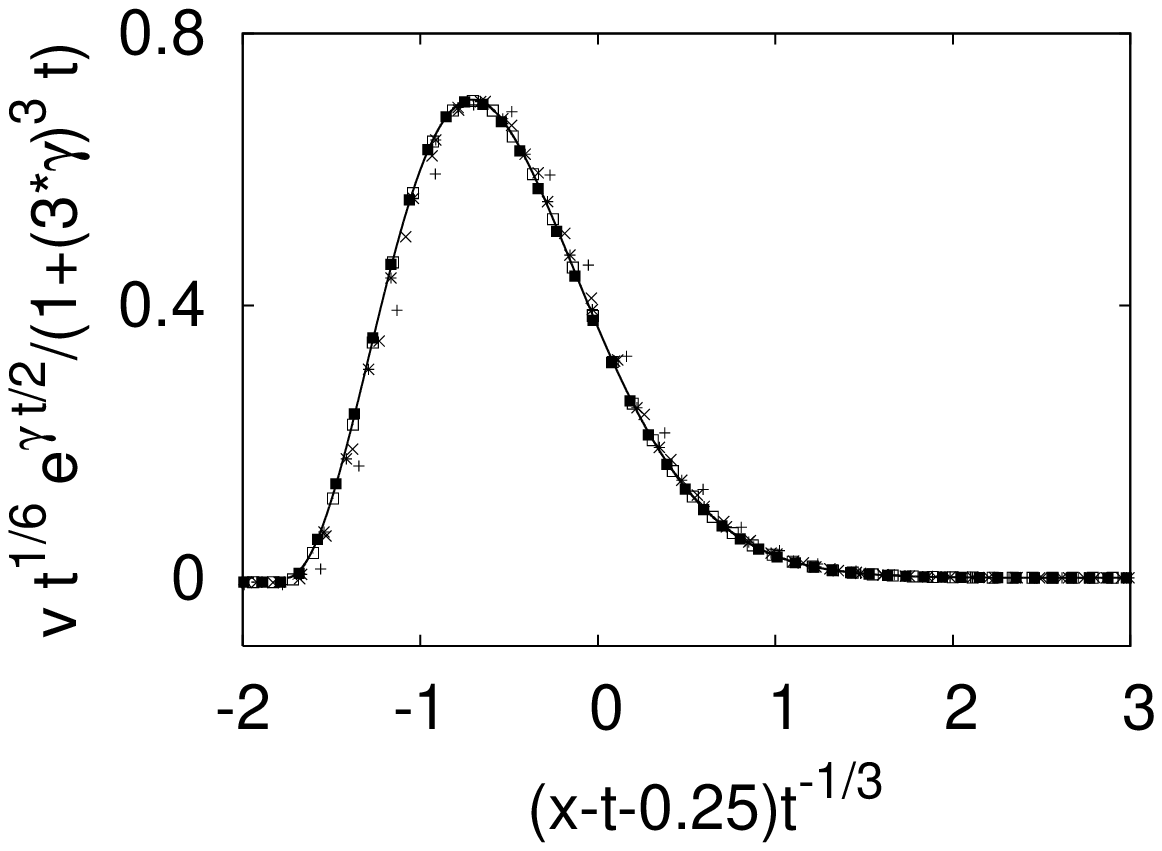}}
\subfigure[$ \gamma = 0.010 $]{\includegraphics[width=6cm]{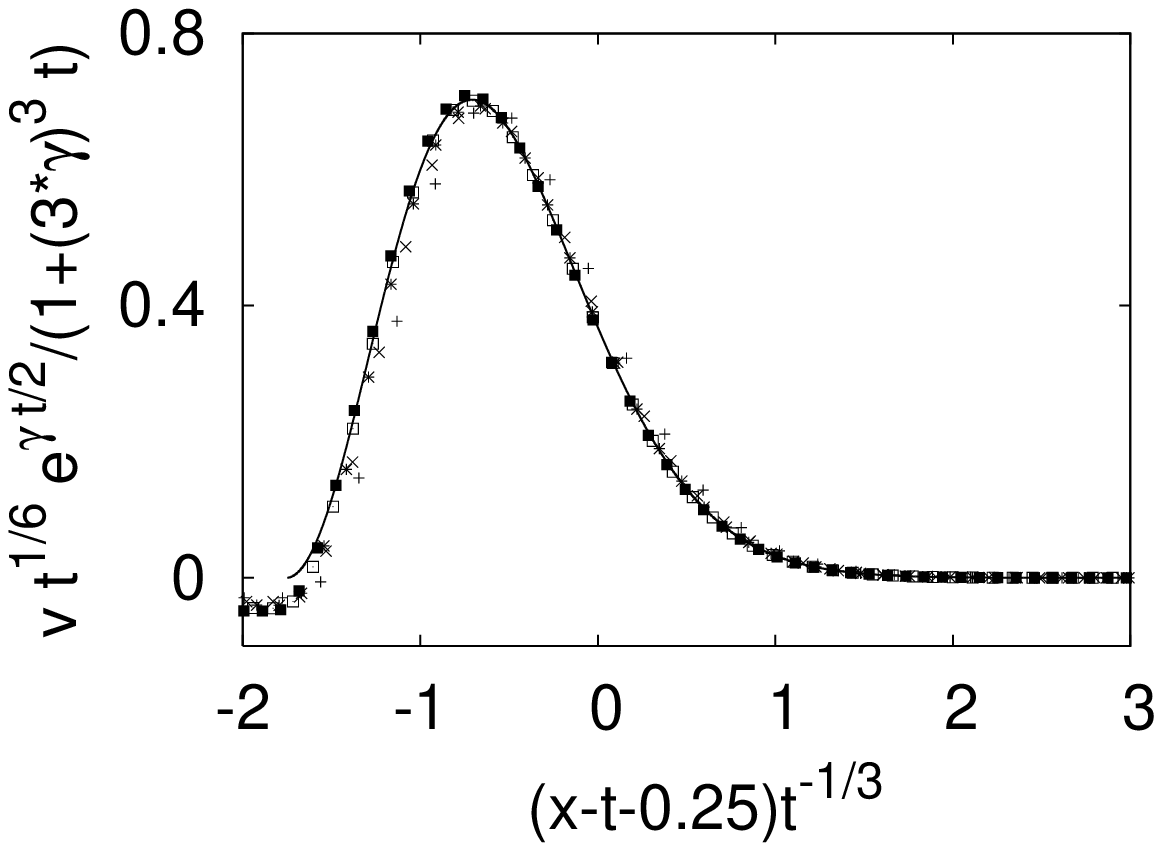}}
\subfigure[$ \gamma = 0.020 $]{\includegraphics[width=6cm]{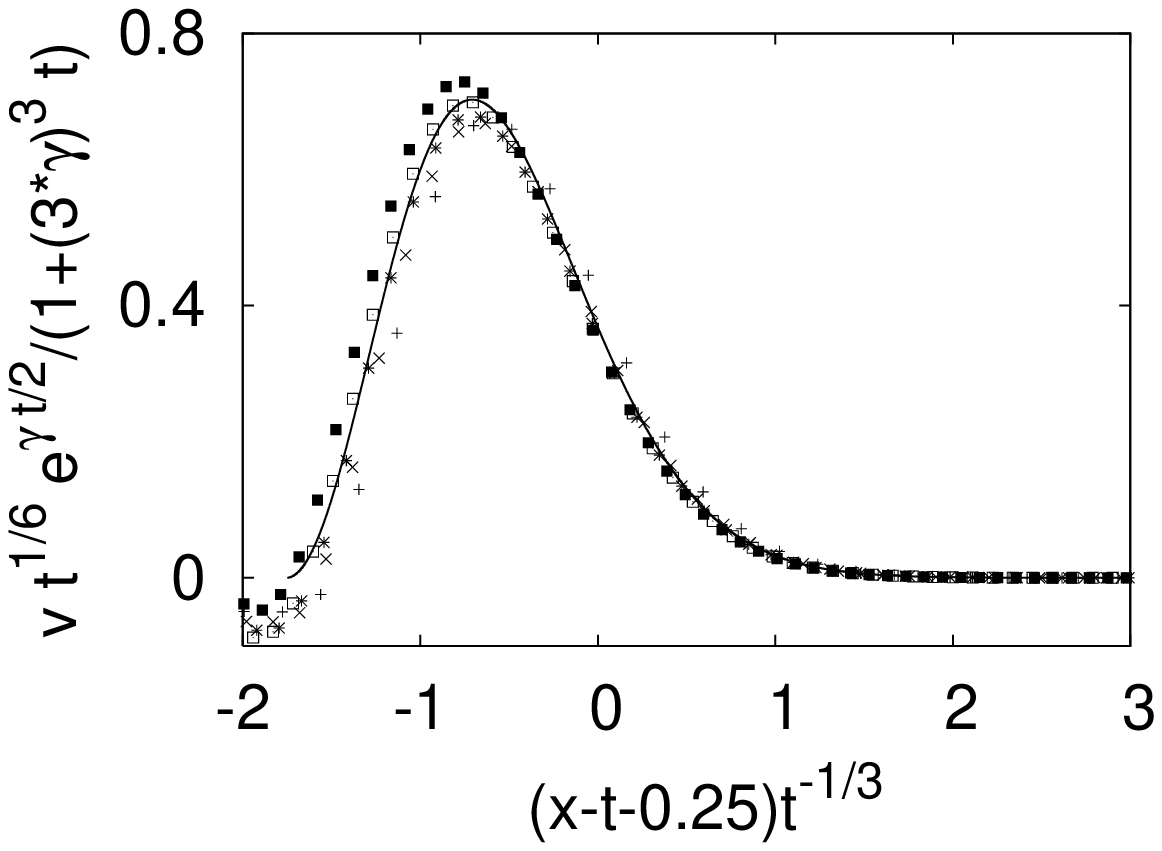}}
\end{center}
\caption{Chain of cylindrical granules: velocity pulse corrected for
higher order frictional effects as
a function of scaled poisition for different friction coefficients.
\label{fig:newscaling}} 
\end{figure}

We have not calculated further corrections analytically, but a numerical
test is possible.  One might test a correction
of $\mathcal{O}(\gamma)$ [cf. Eq.~(\ref{eq:velocityf})], e.g. of the form
$b(t) = t^{1/6} \left [1 + \gamma C(\nu,t)\right ]$, or of
$\mathcal{O}(\gamma^2)$ [cf. Eq.~(\ref{eq:besself})].  The coefficient $C$
in the correction would in general be a function of $\nu$.  We have
tested various corrections with the (admittedly unjustified)
assumption that $C$ does not depend on $\nu$ (the numerical
effort to do otherwise seems unwarranted), and have found that the
leading correction seems to be of $\mathcal{O}(\gamma^3)$, specifically
$b(t) = t^{1/6} (1 + 27\gamma^3 t)$.  This result is analytically
appealing but is purely a numerical outcome. The rescaled velocity
results are shown in Fig.~\ref{fig:newscaling}.

\begin{figure}
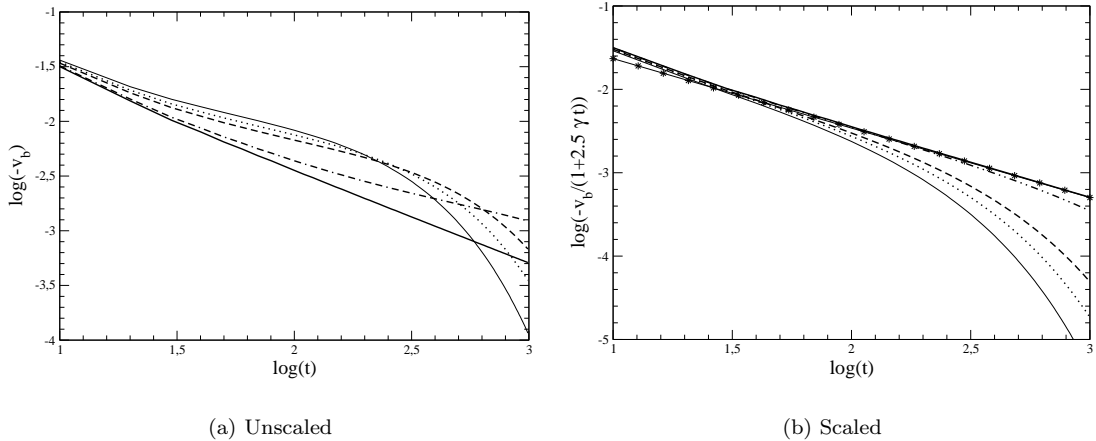

\begin{center}
\subfigure[Unscaled]{\includegraphics[width=7cm]{fig4a.eps}}
\hspace*{0.1in}
\subfigure[Scaled]{\includegraphics[width=7cm]{fig4b.eps}}
\end{center}
\caption{Backward velocity of the last ejected granule as a function of
time. The curves are for $\gamma=0$ (thick line), $0.001$
(dashed-dotted line), $0.005$ (dashed line), $0.007$ (dotted line), and
$0.01$ (thin line).
The line with the star symbols in the second panel has slope $-5/6$.
The slope of the frictionless curve is exactly $-5/6$, as predicted by HSJ.
\label{fig:backscaling}} 
\end{figure}

Finally, we have predicted a somewhat
unexpected feature of the backscattered granules, namely, that at the
moment of ejection their velocity is {\em greater} than that of the
corresponding ejected granules in the absence of friction
[cf. Eq.~(\ref{eq:larger})].  This is seen in
Fig.~\ref{fig:backscaling}.  In the
first panel we show the velocity of the last ejected granule as a function
of time for different values of the friction coefficient.  The velocity is
indeed higher with increasing friction, up to a time beyond which more complex
behavior not captured by our lowest order theory sets in.  In the second
panel we have scaled the velocities by the factor $(1+\frac{5}{2}\gamma t)$
which, according to our theory, should collapse the curves.  At very
early times, while the collapse occurs with the scaling predicted by
our theory, the resulting curve is not yet the theoretical one because
the pulse is not yet clearly defined. However, after this early period
we see both the collapse as well as agreement with the theoretical line
up to times at which higher order effects set in, indicating that
our theory captures the correct behavior up to those times.

%4a,b

\section{Spheres ($n=5/2$)
\label{sec:sphere} }
The archetypal system for the study of impulse dynamics associated
with a Hertz potential of exponent $n>2$ consists of spherical granules.
Such a nonlinear
potential gives rise to a traveling pulse of essentially constant speed
that seems to retain its shape for very long times, i.e. it travels
with essentially constant amplitude and constant width. 
We say ``essentially'' because this is the outcome of approximate
theories~\cite{nesterenko,hinch} and of numerical simulations~\cite{hinch}
but is not a rigorous result (except for $n\to\infty$).  The pulse
arises from the balance of
the dispersive forces that tend to spread the excitation, and the nonlinear
and discrete nature of the system that focuses
it.  As we saw in the last section, discreteness is not sufficient to stop
the widening of the pulse, but only to slow it down.  Nonlinearity
is necessary to obtain a pulse of constant width, 
at the very least approximately.
The result is a narrow pulse, involving no more than a handful of
particles at any time.  The pulse is narrower with increasing $n$, but it
is already sharply defined in the classic case $n=5/2$.  This
sort of behavior is also known from the classic FPU problem where the
potential typically considered is quartic, $n=4$ (or a combination
of $n=4$ and
lower order contributions), and includes restoring forces~\cite{ourpulse}.
The absence of dispersive
restoring forces in a Hertz system causes the localized excitation to be
even more stable than in the associated FPU system.

In this section our analysis again consists of three parts.  We begin by
briefly reviewing the theory for a chain of frictionless spherical granules as
first presented by Nesterenko~\cite{nesterenko}. 
Then we modify this analysis to include
the effects of friction.  In this case there is no exactly solvable
discrete counterpart even if one includes restoring forces.  Finally,
we present numerical results to support our findings. 

\subsection{Frictionless Granules - Theory}
The theory first introduced by Nesterenko~\cite{nesterenko} and later
augmented and complemented by others is based on the same approximation
implemented in the case of cylindrical granules, i.e., a Taylor
expansion of $x_{k\pm 1}$ about $x_k$ and retention of a term beyond the
first (which here is no longer purely diffusive coupling).  This
continuum equation,
assumed to hold within the compression pulse, again does not explicitly
exclude restoring forces and is therefore equivalent to a continuum
approximation with some contributions due to discreteness for the
FPU problem.  The starting equation is thus 
\begin{equation}
\ddot{x}_k =  -(x_k - x_{k+1})^{n-1}  +  (x_{k-1} - x_{k})^{n-1}.
\end{equation}
and the resulting continuum approximation is 
\begin{equation}
\frac{\partial^2 x}{\partial t^2} = \frac{\partial}{\partial k}
\left [- \left ( - \frac{\partial x}{\partial k}\right )^{n-1} +
\frac{n-1}{24} \left ( - \frac{\partial x}{\partial k}\right )^{n-2}
\frac{\partial^3 x}{\partial k^3} \right ] -
\frac{1}{24}\frac{\partial^3}{\partial k^3}
\left [\left ( - \frac{\partial x}{\partial k}\right )^{n-1} \right ].
\label{eq:nestwave}
\end{equation} 
This equation reproduces Eq.~(\ref{eq:wavehinch}) when $n=2$.

As before, one implements a change of variables to a moving frame,
with $\xi=k-c_0t$. Here $c_0$ is a speed to be determined.  The big
difference between the equation one obtains with $n>2$ and that obtained
earlier for $n=2$ [Eq.~(\ref{eq:wavehinchmove})] is that it only
involves the variable $\xi$; the variable $t$ no longer appears explicitly:
\begin{equation}
\frac{\partial }{\partial \xi} \left [ c_0^2
\left (-\frac{\partial x}{\partial \xi}\right ) -
\left ( - \frac{\partial x}{\partial \xi}\right )^{n-1} -
\frac{n-1}{24} \left ( - \frac{\partial x}{\partial \xi}\right )^{n-2}
\frac{\partial^2 }{\partial \xi^2} \left (-\frac{\partial x}
{\partial \xi}\right ) - \frac{1}{24}\frac{\partial^2}{\partial \xi^2}
\left ( - \frac{\partial x}{\partial \xi}\right )^{n-1} \right ] = 0.
\label{eq:movingsph}
\end{equation} 

Nesterenko recognized that there is a simple 
solution to this nonlinear problem:
\begin{equation}
\left (-\frac{\partial x}{\partial \xi}\right ) =
A_0 \sin^m \alpha \xi,
\label{eq:soliton}
\end{equation} 
where $ A_0, $ $ m $ and $\alpha$ are constants.
Substitution of this solution into the propagating equation leads to the
following values:
\begin{equation}
m=\frac{2}{n-2}, \qquad
\alpha=\left(\frac{6(n-2)^2}{n(n-1)}\right)^{1/2},\qquad c_0=(2/n)^{1/2}
A_0^{\frac{n-2}{2}}.
\label{eq:constants}
\end{equation}
An additional assumption is
introduced at this point: a solitary wave is ``constructed'' by
retaining this solution over one period,
$0\leq \alpha(k-c_0t)\leq \pi$, and setting $\partial x/\partial \xi$ equal
to zero outside of this range.
Note that this solution does not satisfy the velocity pulse initial
condition, but rather it describes the solution that the system
presumably settles into after a short initial transient, an
assumption that is supported by numerical simulation results~\cite{hinch}.
For spherical granules
\begin{equation}
\left (-\frac{\partial x}{\partial k}\right ) =
-\left (\frac{5c_0^2}{4}\right )^2 \sin^4 \sqrt{\frac{2}{5}} (k - c_0 t).
\end{equation} 

If the initial velocity impulse is unity, then the initial total
energy (all kinetic) is $1/2$.  The solitary wave (\ref{eq:soliton}), which
describes both the potential and kinetic energy of the system once it
settles, is assumed to contain essentially all of this initial energy
(the numerical simulation results confirm that the energy of the solitary
wave is $99.7\%$ of the initial energy~\cite{hinch}).  Because the
potential is nonlinear, the potential and kinetic energies are
no longer
equal, but one can use the generalized equipartition theorem~\cite{tolman}
to calculate the average contribution of each.  One finds that the ratio is
$K/U=\frac{n}{2}$, so that $K=\frac{n}{2(n+2)}$.
Since the velocity is
\begin{equation}
\dot{x}(\xi)=c_0 A_0 \sin^{\frac{2}{n-2}}\alpha\xi,
\end{equation}
the total kinetic energy in the pulse is 
\begin{equation}
K = \frac{c_0^2 A_0^2}{2\alpha}I\left( \frac{4}{n-2}\right)
=\frac{n}{2(n+2)},
\end{equation} 
where~\cite{gradryz}
\begin{equation}
I(l)\equiv \int_0^\pi \sin^l \theta {\mathrm d}\theta = 
2^{\,l} \; \frac{\Gamma^2\left(\frac{l+1}{2}\right)}{\Gamma(l+1)}.
\label{eq:int}
\end{equation}
The resulting pulse speed $c_0$ and pulse amplitude for $n=5/2$ then are
\begin{equation}
c_0=0.836, \qquad A_0=0.765.
\label{eq:numbers}
\end{equation}
The numerical results of HSJ give $c_0=0.84$.  

Contrary to the $n=2$ case, here there is almost no 
backscattering~\cite{hinch}.  Except for the first two or three granules
that are slightly scattered backwards, the granules in the pulse are
simply displaced by a constant amount and come to rest once the pulse
passes.  The total backward momentum is thus extremely small and finite.
Beyond the first two or three granules, the forward moving
pulse here retains its
shape and amplitude and is therefore essentially conservative with respect to
both energy and momentum.  This is to be contrasted
with the fact that for cylindrical granules, every granule acquires a
backward momentum.

\subsection{Granules With Friction - Theory}
\label{sec:nestgen}

When friction is included, Eq.~(\ref{eq:nestwave}) is modified by the
addition of a frictional term:
\begin{equation}
\frac{\partial^2 x}{\partial t^2} + \gamma \frac{\partial x}{\partial t}
= \frac{\partial}{\partial k} \left [- \left ( - \frac{\partial x}
{\partial k}\right )^{n-1} + \frac{n-1}{24}
\left ( - \frac{\partial x}{\partial k}\right )^{n-2}
\frac{\partial^3 x}{\partial k^3} \right ] - \frac{1}{24}
\frac{\partial^3}{\partial k^3} \left [\left (
- \frac{\partial x}{\partial k}\right )^{n-1} \right ].
\label{eq:nestwavegen}
\end{equation} 
For small values of $ \gamma, $ we expect the
solution~(\ref{eq:soliton}) to be modified in two ways.  Firstly, as we did
with the cylindrical granules, we must take into account the
overall decay of the energy of the pulse, being mindful of the fact that
kinetic and potential energies are not equal when $n>2$.  Secondly,
since the speed of the pulse depends on its amplitude (and hence on its
total energy), we must include the fact that the pulse speed decreases with
time.
We assume a solution of the form
\begin{equation}
\left (-\frac{\partial x}{\partial \xi}\right ) =
A(t) \sin^{\frac{2}{n-2}} \alpha \xi(t),
\label{eq:with}
\end{equation} 
where
\begin{equation}
\xi(k,t) = k - \int_0^t c(t) \mathrm{d} t
\label{eq:xiprime}
\end{equation}
and
\begin{equation}
c(t) = \sqrt{\frac{2}{n}} A^{\frac{n-2}{2}}(t). 
\end{equation}
Note that this form supposes that the width $\pi/\alpha$
of the pulse is not changed by friction~\cite{ourpulse}.

The decay of $A(t)$ [and hence of $c(t)$] can be determined by assuming 
that the pulse energy decays as
$e^{-2u\gamma t}$ and choosing the constant $u$ so that
Eq.~(\ref{eq:nestwavegen}) is satisfied to first order in $\gamma$.
If the pulse energy decays as $e^{-2u\gamma t}$, then the pulse velocity
decays as $e^{-u\gamma t}$. 
For the pulse velocity we have, aside from
its $\xi$ dependence,
\begin{equation}
\frac{\partial x}{\partial t} \sim c(t)A(t) \sim A^{n/2} \sim
e^{-u\gamma t}
\end{equation}
from which it follows that
\begin{equation}
A(t)= A_0 e^{-\frac{2u}{n}\gamma t}, \qquad c(t)=c_0
e^{-\frac{(n-2)u}{n}\gamma t}
\label{eq:A}
\end{equation}
and therefore
\begin{equation}
\xi(k,t) = k - c_0 \frac{n}{u\gamma (n-2)} \left (1 -
e^{-\frac{(n-2)u}{n}\gamma t} \right ).
\label{eq:xinest}
\end{equation} 
To determine $u$ we note that Eq.~(\ref{eq:with}) implies that
$x(t,\xi)=A(t)F(\xi)$, where the form of $F$ is unimportant for the
moment except that it depends only on $\xi$ and not separately on $t$.
Therefore
\begin{equation}
\frac{\partial x}{\partial t}
= \frac{{\mathrm d} A}{{\mathrm d}t}F
+ A\left( \frac{\partial \xi}{\partial t} \right)
\left( \frac{{\mathrm d} F}{{\mathrm d}\xi}\right)
= \left( \frac{1}{A} \frac{{\mathrm d} A}{{\mathrm d}t}\right) x
+ \left( \frac{\partial \xi}{\partial t}\right)
\left(\frac{\partial x}{\partial \xi}\right).
\label{eq:mid1}
\end{equation}
Similarly we find
\begin{equation}
\frac{\partial^2 x}{\partial x^2}
= \left( \frac{1}{A} \frac{{\mathrm d}^2 A}{{\mathrm d}t^2}\right) x
+ 2\left( \frac{1}{A} \frac{{\mathrm d} A}{{\mathrm d}t}\right)
\left( \frac{ \partial \xi}{\partial t}\right)
\left( \frac{\partial x}{\partial \xi}\right) + \left( \frac{\partial^2
\xi}{\partial t^2}\right)\left(\frac{\partial x}{\partial \xi}\right)
+ \left( \frac{\partial \xi}{\partial t} \right)^2
\left(\frac{\partial^2 x}{\partial \xi^2}\right).
\label{eq:mid2}
\end{equation}
Substitution of Eqs.~(\ref{eq:A}) and (\ref{eq:xinest}) into
(\ref{eq:mid1}) and (\ref{eq:mid2}) gives
\begin{equation}
\frac{\partial^2 x}{\partial t^2} + \gamma \frac{\partial x}{\partial t}
= c_0^2 e^{-\frac{2(n-2)u}{n}\gamma t}\frac{\partial^2 x}{\partial
\xi'^2} + \gamma\left(\frac{4u}{n} +\frac{(n-2)u}{n}-1 \right)
c_0 e^{-\frac{(n-2)u}{n}\gamma t} +\mathcal{O} (\gamma^2).
\end{equation}
One must then choose
\begin{equation}
u=\frac{n}{n+2}
\end{equation}
to force the $\mathcal{O}(\gamma)$ contribution to vanish.  We are then
left with
\begin{equation}
\frac{\partial^2 x}{\partial t^2} + \gamma \frac{\partial x}{\partial t}
= c_0^2 e^{-\frac{2(n-2)}{(n+2)}\gamma t}\frac{\partial^2 x}{\partial
\xi'^2} +\mathcal{O} (\gamma^2).
\end{equation}
For $\gamma t\ll (n+2)/2(n-2)$, substitution 
into Eq.~(\ref{eq:nestwavegen}) leads again to
Eq.(\ref{eq:movingsph}) to $\mathcal{O} (\gamma^2)$.

We thus conclude that to $\mathcal{O} (\gamma^2)$ and for times $\gamma
t\ll (n+2)/2(n-2)$ (which means essentially the entire lifetime of the
pulse, see next subsection) the solution for the chain of spherical
granules subject to weak friction is as assumed in Eqs.~(\ref{eq:with}) and
(\ref{eq:xiprime}) with
\begin{equation}
A(t)=A_0 e^{-\frac{2}{n+2}\gamma t}, \qquad c(t)=\sqrt{\frac{2}{n}}
A_0 ^ {\frac{n-2}{2}} e^{-\frac{(n-2)}{(n+2)}\gamma t}.
\label{eq:timedep}
\end{equation}
The shape of the pulse is constant and
the same as in the frictionless case.
The width remains constant in time, the overall energy in the pulse
decays as $e^{-\frac{2n}{n+2}\gamma t}$, and the pulse velocity as
$e^{-\frac{n}{n+2}\gamma t}$.  For $n=5/2$ these decays go as
$e^{-(10/9)\gamma t}$ and $e^{-(5/9)\gamma t}$ respectively.
It is an interesting sideline to note the increase in the
friction-induced decay
rate of the velocity or energy of the compression pulse with increasing $n$;
for $n\rightarrow \infty$ it is the same as that of a single particle
traveling in a viscous medium.

%5
\begin{figure}
\begin{center}
\includegraphics[width=8cm]{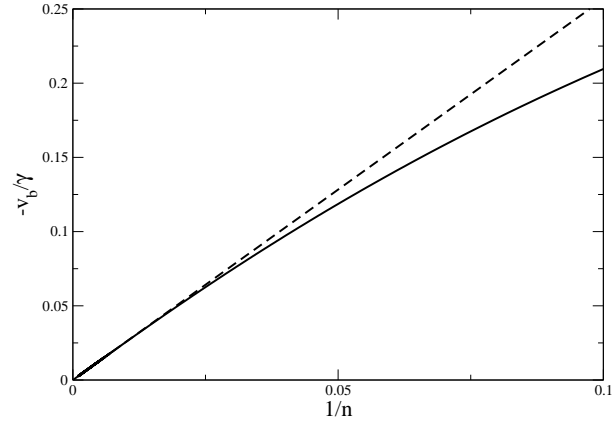}
\caption{
Backscattering velocity at the moment of ejection as a function of the
power $n$ in the Hertz potential.  The dashed line is the 
asymptote Eq.~(\ref{eq:vb}), which in this representation is
independent of $\gamma$ and
of $t$.  The dark curve is the full result Eq.~(\ref{eq:vbhertz})
for $\gamma t=1$.  
\label{fig:vb}}
\end{center}
\end{figure} 

%6
\begin{figure}
\begin{center}
\includegraphics[width=8cm]{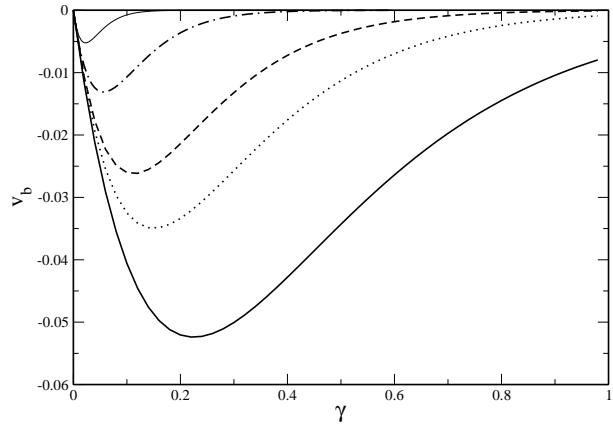}
\caption{Backscattering velocity at the moment of ejection
as a function
of the friction parameter for spherical granules at different times
as follows: $t=10$ (thick), $15$ (dotted), $20$ (dashed), $40$
(dot-dashed), and $100$ (thin).
\label{fig:vbgamma}}
\end{center}
\end{figure} 

%7
\begin{figure}
\begin{center}
\includegraphics[width=8cm]{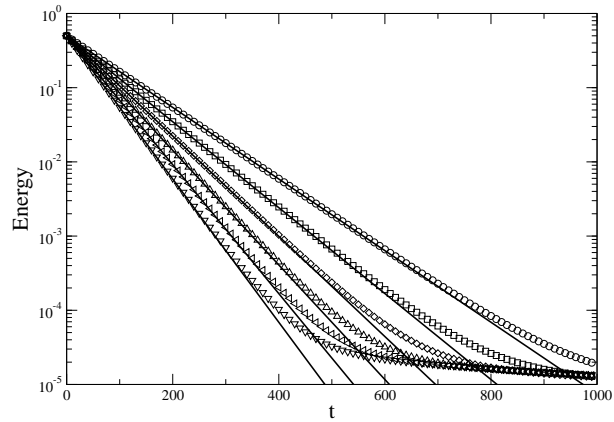}
\caption{
Total energy decay as a function of time. Within a time range of
$\mathcal{O}(10\gamma^{-1})$ the energy decays
exponentially as $e^{-(10/9)\gamma t}$, drawn as lines.  The symbols
are numerical simulation results. From top to bottom the curves
are for $\gamma = 0.010, 0.012, 0.014, 0.016, 0.018$, and $0.020$.
\label{fig:energyhertz}}
\end{center}
\end{figure} 

A remarkable difference between the frictionless and frictional chains
of spherical granules lies in the backward scattering.  In the
frictionless case we noted that there is almost no backward scattering,
and the very small amount of it appears only in the first three or so
particles. With friction, {\em each} particle is scattered backward as
the pulse leaves, more like the situation in the cylindrical granule
case.  Following the reasoning implemented in that case we first
calculate the forward momentum of the pulse,
\begin{eqnarray}
P &=& \sum_{\dot{x}(k,t)>0} \dot{x}(k,t) \simeq \sum_{\dot{x}(k,t)>0}
\left( \frac{\partial x}{\partial \xi }\right)
\left( \frac{\partial
\xi}{\partial t}\right) \nonumber\\
&=& \sum_{\dot{x}(k,t)>0} A(t) c(t) \sin^{\frac{2}{n-2}} \alpha\xi
= \frac{A(t)c(t)}{\alpha} I\left(\frac{2}{n-2}\right)\nonumber\\
&=&
\frac{(2/n)^{1/2}}{\alpha} A_0^{\frac{n}{2}} I\left(\frac{2}{n-2}\right)
e^{-\frac{n}{n+2}\gamma t},
\end{eqnarray} 
where $I(l)$ is given in Eq.~(\ref{eq:int}).
The rate of change of the forward momentum then is
\begin{equation}
\dot{P} =
-\gamma \frac{(2/n)^{1/2}}{\alpha} \frac{n}{(n+2)}
A_0^{\frac{n}{2}} I\left(\frac{2}{n-2}\right)
e^{-\frac{n}{n+2}\gamma t}
\end{equation}
On the other hand, the loss of momentum of the last particle in
the pulse as the pulse
moves across it is [cf. Eq.~(\ref{eq:loss})]
\begin{eqnarray}
\delta & =& -\gamma\Delta x =-\gamma\frac{A(t)}{\alpha}
I\left(\frac{2}{n-2}\right)\nonumber\\
&=& -\gamma\frac{A_0}{\alpha}I\left(\frac{2}{n-2}\right) 
e^{-\frac{2}{n+2}\gamma t}.
\end{eqnarray}
The momentum transferred to the last particle as it is ejected from the
pulse is therefore
\begin{equation}
v_b = -\frac{\dot{P}}{c(t)} +\delta = 
-2\gamma \frac{A_0}{\alpha(n+2)}I\left(\frac{2}{n-2}\right)
e^{-\frac{2}{n+2}\gamma t}
\label{eq:sphback}
\end{equation}
(in Eqs.~(\ref{eq:vsubb}) and (\ref{eq:vfsubb}) the factor $c(t)$ does
not appear explicitly because it is equal to unity).
For spherical granules 
\begin{equation}
v_b=- \gamma\sqrt{\frac{5}{2}}~\frac {\pi A_0}{6}e^{-\frac{4}{9}\gamma t}.
\label{eq:vbhertz}
\end{equation}
Every particle in the chain is therefore ejected with a (exponentially
decreasing) backward
momentum, in contrast to the frictionless chain.  The loss of energy in
the pulse due to friction is not balanced by a sufficiently large loss
of momentum.  This relative increase of pulse momentum must be balanced
by the momentum carried by the ejected granules. 

From Eq.~(\ref{eq:sphback}) we can deduce two additional interesting
results.  One is the dependence of the backward velocity on the power
$n$ of the potential of interaction.  We find that $v_b$ decreases
monotonically with $n$, and for large $n$,
\begin{equation}
v_b = -\frac{\sqrt{\frac{2}{3}}\pi\gamma}{n} +
\mathcal{O}\left(\frac{\log(n)}{n^2}\right)
\label{eq:vb}
\end{equation}
independently of time to leading order. 
The approach to this asymptotic result is shown in
Fig.~\ref{fig:vb}, where we present $v_b/\gamma$ as a function of $1/n$.
The other is the nonmonotonic dependence of $v_b$ on the damping, a
result already found in our two-granule analysis~\cite{we} and
illustrated in Fig.~\ref{fig:vbgamma}.  For spherical granules we show
this dependence of $v_b$ on damping at different times.  This
qualitative behavior persists for other values of $n>2$: as $n$
increases the position of the minimum is shifted to larger $\gamma$ 
and the absolute value of the minimum decreases.

Equation~(\ref{eq:with}) with (\ref{eq:timedep}) and
Eq.~(\ref{eq:sphback}) are the principal results of this subsection.  
They establish analytic expressions for the pulse and for the backscattered
momentum that will be checked against numerical simulations in the
following subsection.

\subsection{Numerical Simulations}
First we note that even in the frictionless case the pulse velocity
is lower here
($c_0=0.836$) than in the cylindrical case (unit velocity), a result
obtained in the numerical simulations of HSJ as well as in our theory, cf.
Eq.~(\ref{eq:numbers}).  With friction, the pulse speed decreases as the
pulse loses energy.

As with cylindrical granules,
for times $t\lesssim 10\gamma^{-1}$ the pulse
has the same shape in the presence of friction as in the frictionless case
except for the overall exponential decay, as
illustrated in Fig.~\ref{fig:energyhertz}.
Note that the decay is more rapid than in the case of cylindrical granules. 

\begin{figure}
\begin{center}
\includegraphics[width=8cm]{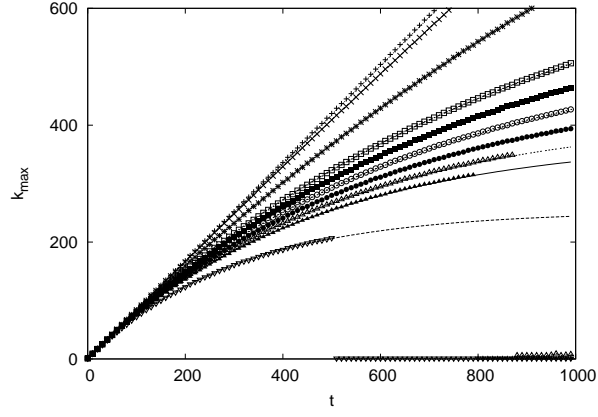}
\caption{Grain with the maximum velocity as a function of time.  The
symbols are simulation results, and the lines are Eq.~(\ref{eq:kmax}).  
From top to bottom $\gamma = 0, 0.001, 0.005, 0.01, 0.012, 0.014, 0.016,
0.018, 0.02, 0.03$.  The $\gamma=0$ curve is $k_max=(\pi/2\alpha)+c_0t$.
\label{fig:kmaxhertz}}
\end{center}
\end{figure}

Contrary to the cylindrical granule case, the pulse also
slows down as its energy decreases.  This behavior continues
until the energy begins to decrease abruptly (more rapidly than
exponentially) at a time $\sim 8.3\gamma^{-1}$.  At a time $\sim
16.2\gamma^{-1}$ the backscattered energy becomes greater than the pulse
energy and at the same time the pulse stops moving.  In
Fig.~\ref{fig:kmaxhertz} we show $k_{max}$, the granules with the maximum
velocity in the pulse.  The symbols are the results
obtained from numerical simulations, and the lines are 
[cf. Eq.~(\ref{eq:xinest})]
\begin{equation}
k_{max}=\frac{\pi}{2\alpha} +\frac{c_0}{\gamma} \frac{(n+2)}{(n-2)} \left(
1-e^{-\frac{(n-2)}{(n+2)}\gamma t}\right) =
\left(\frac{5}{2}\right)^{1/2}\frac{\pi}{2} + \frac{9c_0}{\gamma} \left(
1-e^{-\gamma t/9}\right),
\label{eq:kmax}
\end{equation}
where the last expression holds explicitly for $n=5/2$.  The value used for
$c_0$ is as given in Eq.~(\ref{eq:numbers}).  The agreement is clearly
excellent.

%8

%9a,b,c,d
\begin{figure}
\begin{center}
\subfigure[$ \gamma = 0.000 $]{\includegraphics[width=6cm]{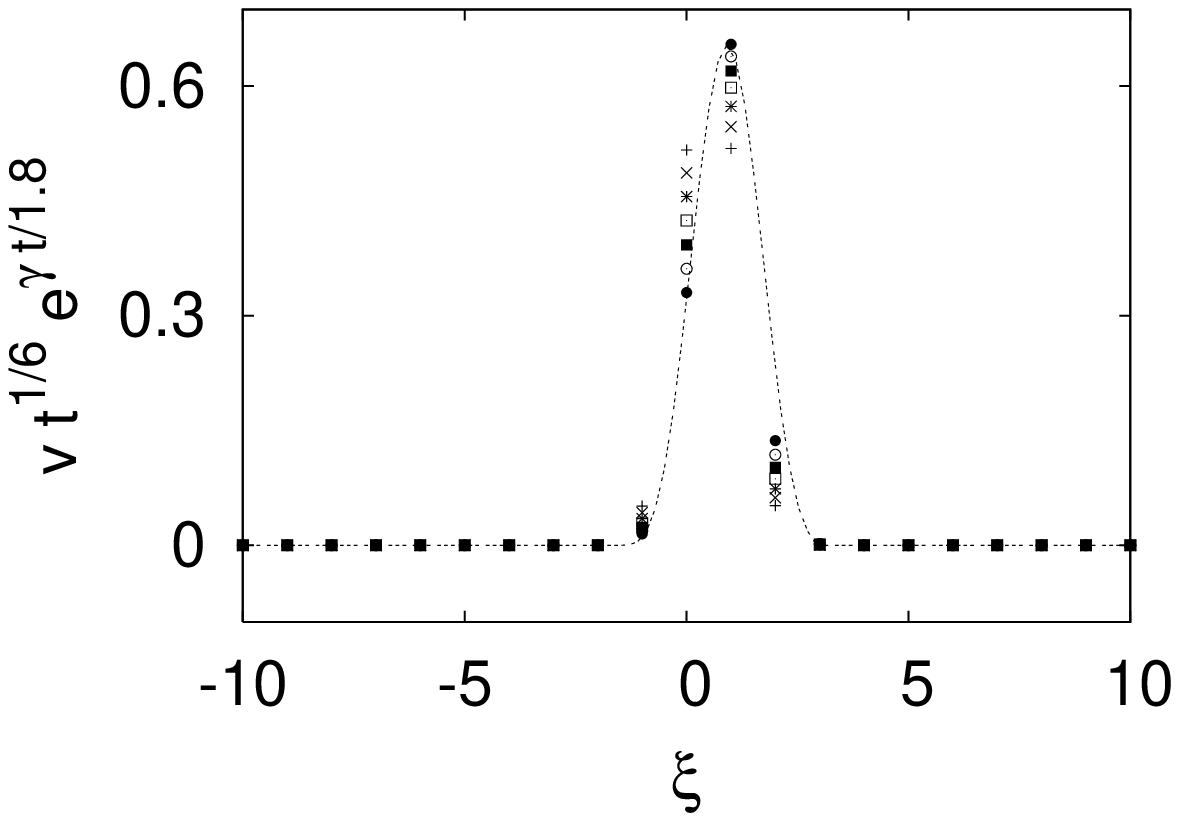}}
\subfigure[$ \gamma = 0.010 $]{\includegraphics[width=6cm]{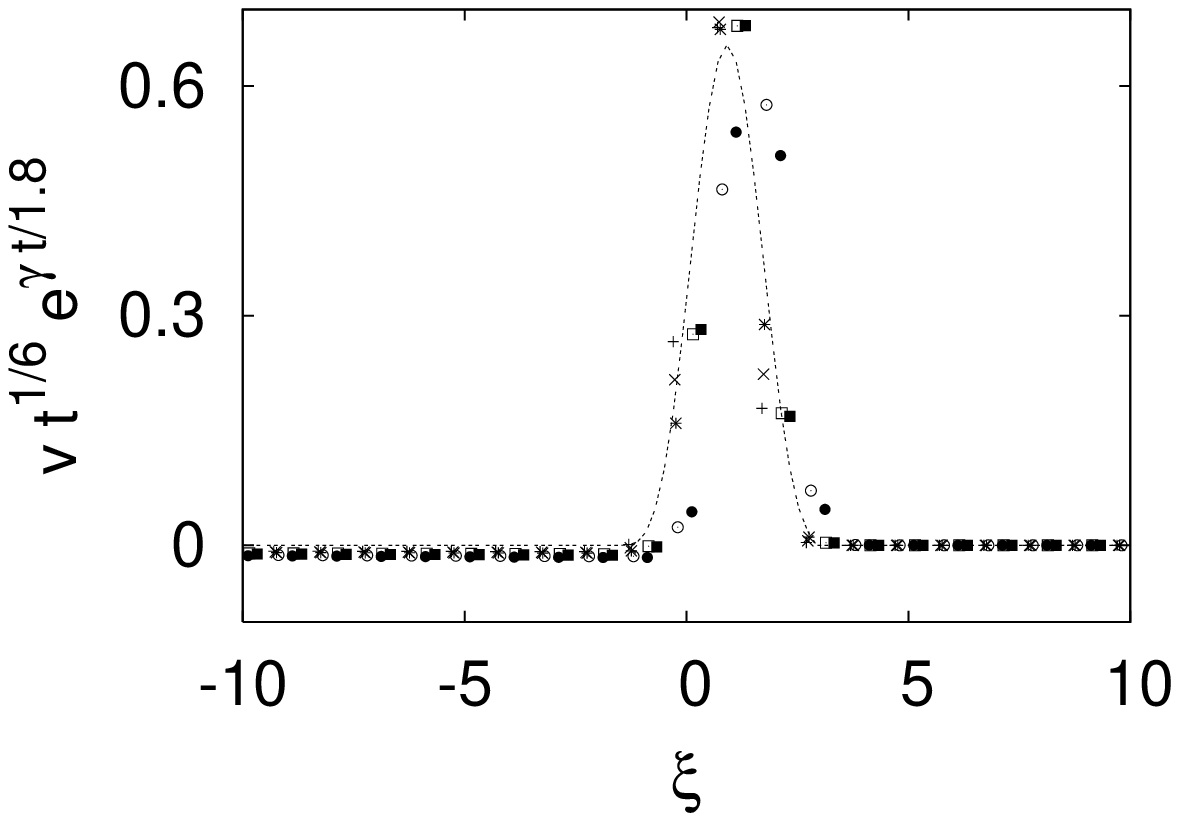}}
\subfigure[$ \gamma = 0.016 $]{\includegraphics[width=6cm]{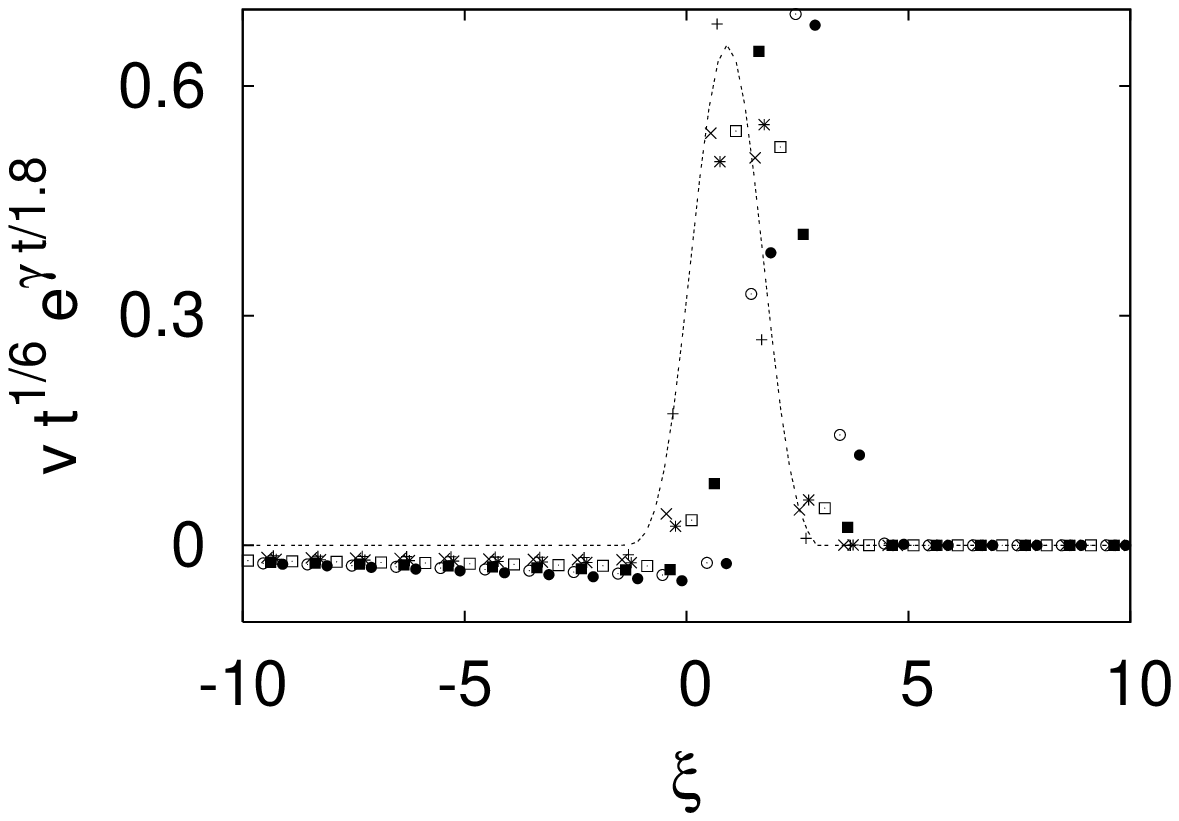}}
\subfigure[$ \gamma = 0.020 $]{\includegraphics[width=6cm]{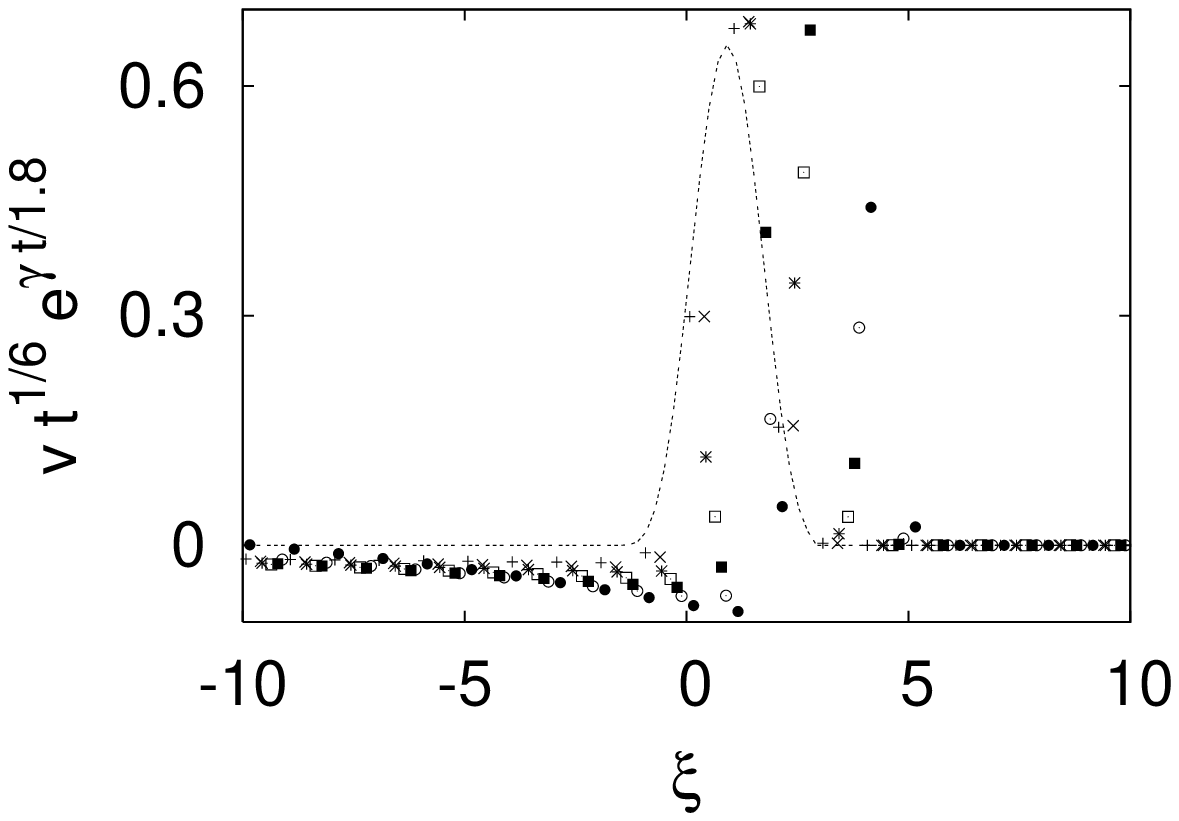}}
\end{center}
\caption{Scaled velocity pulse as a function of scaled position for
different friction coefficients.
\label{fig:nestscaling}} 
\end{figure}

\begin{figure}
\begin{center}
\includegraphics[width=8cm]{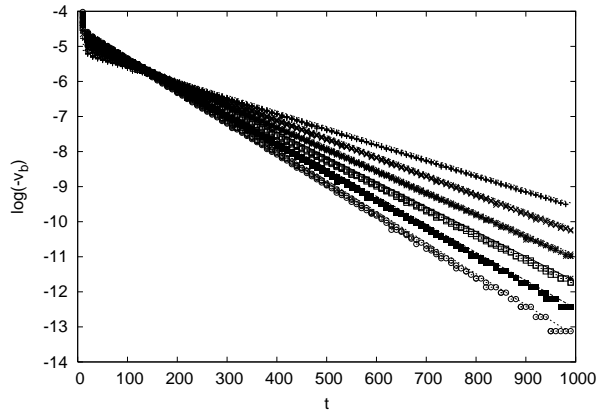}
\caption{Backscattering velocity at the moment of ejection as a function of
time for different friction coefficients.  The symbols are simulation
results and the lines are Eq.~(\ref{eq:vbhertz}).  
From top to bottom on the right side of the figure
$\gamma =  0.010, 0.012, 0.014, 0.016, 0.018,
0.020$.  
\label{fig:back}}
\end{center}
\end{figure}

To check our prediction for the effect of friction on the propagating pulse 
we present the scaled velocity pulse as a function of scaled position for
different friction coefficients in Fig.~\ref{fig:nestscaling}. 
The agreement here is less satisfactory
than in the cylindrical granule case. One difficulty is the fact that
the analytic theory is a continuum approximation while the chain is
discrete, and here only a very small number of granules are actually
moving forward at any one time, that is, the pulse is very narrow.  A
second difficulty is that in addition to the smooth envelope that the
continuum theory attempts to capture, the narrow pulse actually 
experiences small amplitude oscillations as it moves forward.  Our reported
values in Fig.~\ref{fig:nestscaling} include values that might fall anywhere
within these oscillations.  It is nevertheless clear that the
theory captures the qualitative features of the pulse.  In particular, we
point to the increasing backscattering with increasing friction
that can be seen in these figures.  This is a qualitative difference
between the frictionless chain (where only about three granules backscatter
slightly) to the chain with friction, where some backscattering occurs at
each granule as the pulse passes by.  In Fig.~\ref{fig:back} the symbols
are the simulation results and the lines represent Eq.~(\ref{eq:vbhertz}). 
The agreement is clearly excellent.

%10

\section{Conclusions}
\label{sec:conclusion}
In this work we have studied the dynamics of an initial velocity impulse
in a chain of granules that interact only when in contact, that is, they
experience only a repulsive potential.  Our interest has been in
establishing the effects of hydrodynamic friction on these dynamics.

First we analyzed a chain of cylindrical granules, that is, with a power
law repulsive potential with exponent $n=2$ (half a harmonic potential).
We presented the frictionless case, and organized existing information in a
particular way to clarify the effects of discreteness and of the absence
of restoring forces on these results.  We were also able to obtain a
number of results analytically that had previously only been obtained
numerically~\cite{hinch}. 
In this chain the impulse travels at unit velocity
as a spreading pulse (as $t^{1/3}$) in which the maximum displacement
progressively grows with time (as $t^{1/6}$). While this
traveling pulse carries most of the initial energy, conservation of
momentum requires that there be backscattering of each granule as the
pulse passes by.  The chain thus continually undergoes fragmentation.

We then generalized these results in the presence of friction and found
that the principal effect of weak friction is an overall exponential
decay of the energy.  The pulse still moves at unit velocity, still
spreads as $t^{1/3}$, and the maximum displacement now varies as
$t^{1/3} t^{-\gamma t/2}$ throughout its lifetime. 
There is again backscattering of each granule as the pulse passes by.
An interesting and unexpected effect of
friction is that the velocity of the backscattered particles
at the moment of ejection is {\em greater} than the velocity in the
frictionless chain~\cite{we}.  The backscattered particles also slow
down due to
friction, but this chain, too, undergoes continual fragmentation.
We supported our results via numerical simulations. 

Next we analyzed a chain of granules with a power law repulsive
potential with exponent $n>2$, with special attention to spherical
granules ($n=5/2$).  We reviewed Nesterenko's theory for the
frictionless case, which leads to an essentially conservative pulse of
constant width determined by the power $n$, traveling down the chain
at a velocity that depends on
the energy of the pulse.  This velocity is lower than that of the
spreading pulse in the cylindrical granule case.  Here again we obtained
some results analytically that had previously been reported
numerically~\cite{hinch}.  Contrary to the $n=2$ case there is
essentially no backscattering (fragmentation) in this system: only the
first two or three granules acquire a very small backward momentum as
the pulse passes over.   

The generalization of these results in the presence of friction are more
complicated because an overall decay of the energy 
causes the pulse to slow down as it moves. We found that the solution 
is one in which the overall shape of the pulse as well as its width
remain unchanged, the
energy decay is exponential, as is the decrease in
the displacement pulse amplitude and the pulse velocity. 
The exponential decay factors are fixed by the
power $n$ of the potential.  Most dramatically, we found that in the
presence of friction there is now backscattering of each granule as the
pulse passes by, so that this chain experiences fragmentation.  The
velocity of the backscattered granules of course also decreases
exponentially.  We supported these results for the case of spherical
granules via numerical simulations.

A number of interesting problems immediately come to mind as a possible
extension of this work.  Among them is consideration of the effect of
mass disorder and/or frictional disorder in the chains and of mass
tapering.  Work along these directions is in progress~\cite{future}.

\section*{Acknowledgments}
This work was supported by the Engineering Research Program of
the Office of Basic Energy Sciences at the U. S. Department of Energy
under Grant No. DE-FG03-86ER13606.

\end{document}